\pdfoutput=1
\documentclass[11pt,twoside,a4paper,cmspaper,final,collab]{cms-tdr}
\def\svnVersion{38a24d9-D}\def\svnDate{2020/12/02}\def\cmsCernNoTag{CERN-EP-2020-136}\def\cmsCernDate{\today}\def\cmsMessage{"Published in the European Physical Journal C as \href{http://dx.doi.org/10.1140/epjc/s10052-020-08739-5}{\doi{10.1140/epjc/s10052-020-08739-5}.}"}
\begin{document}\cmsNoteHeader{EXO-19-003}

\newcommand{\mll}{\ensuremath{m_{\ell\ell}}\xspace}
\newcommand{\usedLumi}{137\fbinv}
\newcommand{\DMSIMP} {{\textsc{dmsimp}}\xspace}
\newcommand{\ZZ}{\ensuremath{\PZ\PZ}\xspace}
\newcommand{\WZ}{\ensuremath{\PW\PZ}\xspace}
\newcommand{\WW}{\ensuremath{\PW\PW}\xspace}
\newcommand{\aboson}{\ensuremath{\textsf{a}}\xspace}
\newcommand{\ptvecll}{\ensuremath{\ptvec^{\kern1pt\ell\ell}}\xspace}
\providecommand{\PAGc}{{\HepAntiParticle{\PGc}{}{}}\Xspace}
\newlength\cmsTabSkip\setlength{\cmsTabSkip}{1ex}
\ifthenelse{\boolean{cms@external}}{\providecommand{\cmsLeft}{upper\xspace}}{\providecommand{\cmsLeft}{left\xspace}}
\ifthenelse{\boolean{cms@external}}{\providecommand{\cmsRight}{lower\xspace}}{\providecommand{\cmsRight}{right\xspace}} 

\cmsNoteHeader{EXO-19-003}

\title{Search for dark matter produced in association with a leptonically decaying $\PZ$ boson in proton-proton collisions at $\sqrt{s}=13\TeV$}

\titlerunning{Search for dark matter produced in association with a leptonically decaying $\PZ$ boson}

\author*[inst1]{CMS experiment}

\date{\today}

\abstract{
A search for dark matter particles is performed using events with a \PZ boson candidate and large missing transverse 
momentum.  The analysis is based on proton-proton collision data at a center-of-mass energy of 13\TeV, collected by the CMS 
experiment at the LHC in 2016--2018, corresponding to an integrated luminosity of 137\fbinv. The search uses the decay 
channels $\PZ\to\Pe\Pe$ and $\PZ\to\PGm\PGm$. No significant excess of events is observed over the background expected from 
the standard model. Limits are set on dark matter particle production in the context of simplified models with vector, axial-vector, scalar, 
and pseudoscalar mediators, as well as on a two-Higgs-doublet model with an additional pseudoscalar mediator. In addition, limits are provided
for spin-dependent and spin-independent scattering cross sections and are compared to those from direct-detection experiments. 
The results are also interpreted in the context of models of invisible Higgs boson decays, unparticles, and large extra dimensions.
}

\hypersetup{%
pdfauthor={CMS Collaboration},%
pdftitle={Search for dark matter produced in association with a Z boson in proton-proton collisions at sqrts=13 TeV},%
pdfsubject={CMS},%
pdfkeywords={CMS, physics, mono-Z, dark matter, extra dimensions, unparticles}}

\maketitle

\section{Introduction}\label{sec:introduction}

{\tolerance=800
The existence of dark matter (DM) is well established from astrophysical 
observations~\cite{Bertone:2016nfn}, where
the evidence relies entirely on gravitational interactions. 
According to fits based on the Lambda cold dark matter model of cosmology~\cite{Daniel:2010ky} to observational data, DM comprises 26.4\% of the 
current matter-energy density of the universe, while baryonic matter accounts for only 4.8\%~\cite{Akrami:2018vks}.   
In spite of the abundance of DM, its nature remains unknown. This mystery is the subject of 
an active experimental program to search for dark matter particles, including direct-detection experiments that search for interactions of
ambient DM with ordinary matter, indirect-detection experiments that search for the products of self-annihilation of DM in outer space, 
and searches at accelerators and colliders that attempt to create DM in the laboratory.
\par}

The search presented here considers a ``mono-\PZ'' scenario where a \PZ boson, produced in proton-proton ($\Pp\Pp$) collisions, recoils 
against DM or other beyond the standard model (BSM) invisible particles. 
The \PZ boson subsequently decays into two charged leptons ($\ell^{+}\ell^{-}$, where 
$\ell=\Pe$ or $\PGm$) yielding a dilepton signature, and the accompanying undetected particles contribute to 
missing transverse momentum. The analysis is based on a data set of $\Pp\Pp$ collisions
at a center-of-mass energy of 13\TeV produced at the CERN LHC. The data  were recorded with the CMS detector in 
the years 2016--2018, and 
correspond to an integrated luminosity of $\usedLumi$.  The results are interpreted in the context of 
several models for DM production, as well as for two other scenarios of BSM physics that also predict invisible particles. 

These results extend and supersede a previous search by CMS in the mono-\PZ channel based on a data set collected at $\sqrt{s}=13\TeV$ corresponding to an integrated luminosity of 36\fbinv~\cite{Sirunyan:2017qfc}.  The ATLAS experiment has published searches in this channel as 
well with the latest result based on a data set corresponding to an integrated luminosity of 36\fbinv~\cite{Aaboud:2017bja}.
Similar searches for DM use other ``mono-X" signatures with missing transverse momentum 
recoiling against a hadronic 
jet~\cite{Aaboud:2017phn,Sirunyan:2017jix}, a photon~\cite{Aaboud:2017dor},
 a heavy-flavor (bottom or 
top) quark~\cite{Aaboud:2017rzf,Aad:2020sgw,Sirunyan:2018gka}, a $\PW$ or $\PZ$  boson decaying to 
hadrons~\cite{Sirunyan:2017jix,Aaboud:2017bja,Aaboud:2018xdl},
 or a Higgs 
boson~\cite{Aaboud:2017uak,Aaboud:2017yqz,Sirunyan:2018qob,Sirunyan:2018gdw,Sirunyan:2018fpy,Sirunyan:2019zav}. An additional DM interpretation is explored in searches for Higgs boson decays to invisible particles~\cite{Sirunyan:2018owy,Aaboud:2019rtt}. 

The paper is organized as follows.  The DM and other BSM models explored are introduced along with their relevant parameters in Section~\ref{sec:models}.
Section~\ref{sec:CMS_Detector} gives a brief description of the CMS detector.
 The data and simulated samples are described in Section~\ref{sec:objects}, along with the event reconstruction.  
 The event selection procedures and background estimation methods are described in
 Sections~\ref{sec:selection} and \ref{sec:backgrounds}, respectively. 
Section~\ref{sec:fitting} details the fitting method implemented for the different models presented, while 
Section~\ref{sec:systematics}  discusses the systematic uncertainties. 
The results are given in Section~\ref{sec:results}, and the paper is summarized in Section~\ref{sec:summary}. 

\section{Signal models}\label{sec:models}

Several models of BSM physics can lead to a signature of a \PZ boson subsequently decaying into a lepton pair and missing transverse momentum. The 
goal of this paper is to explore a set of benchmark models for the production of DM that can contribute to this final state.  
In all DM models we consider, the DM particles are produced in pairs,  $\PGc\PAGc$, where  $\PGc$ is 
assumed to be a Dirac fermion. 

\begin{figure*}[!htbp]
  \centering
    {\includegraphics[width=0.44\textwidth]{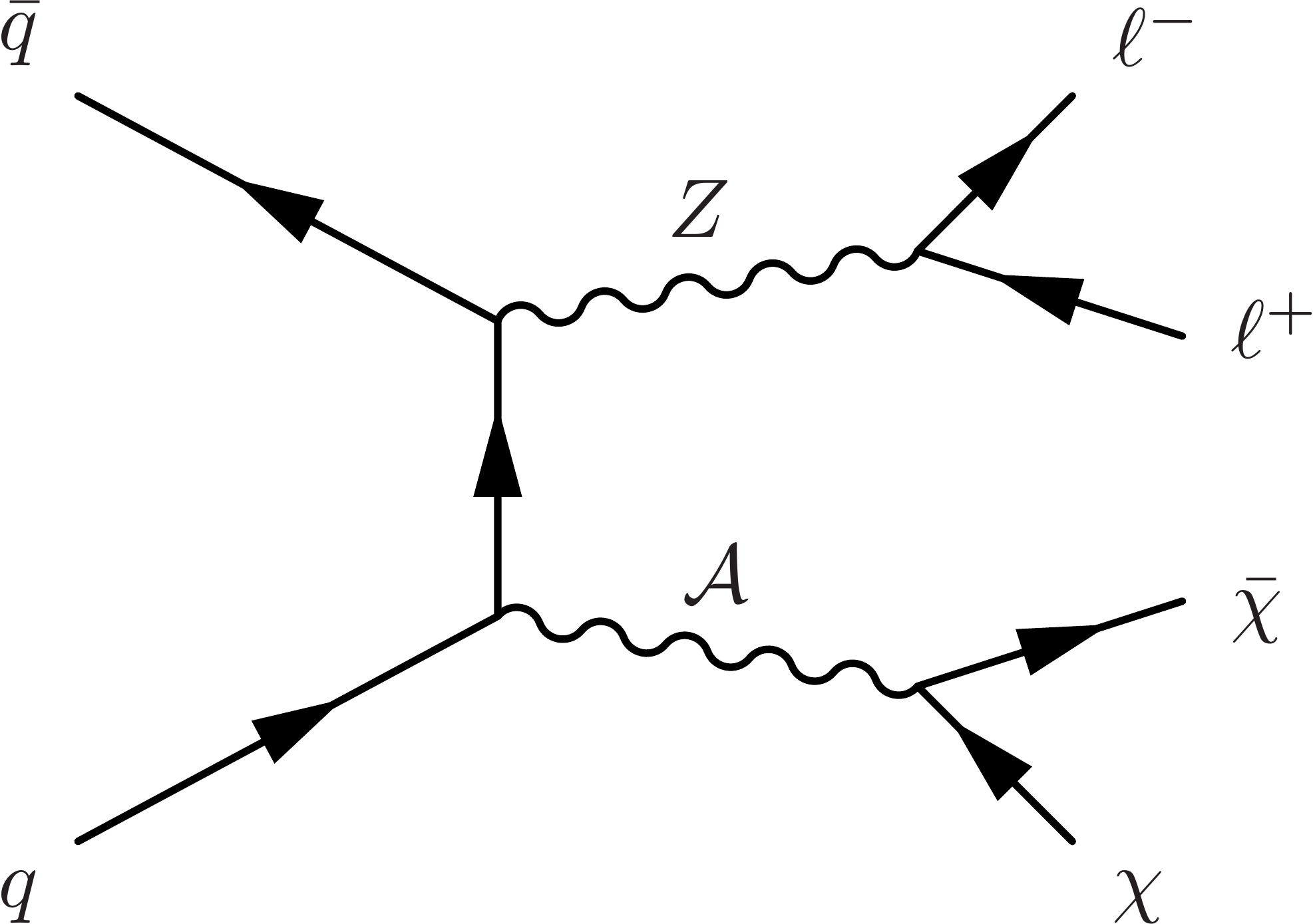}}
    {\includegraphics[width=0.49\textwidth]{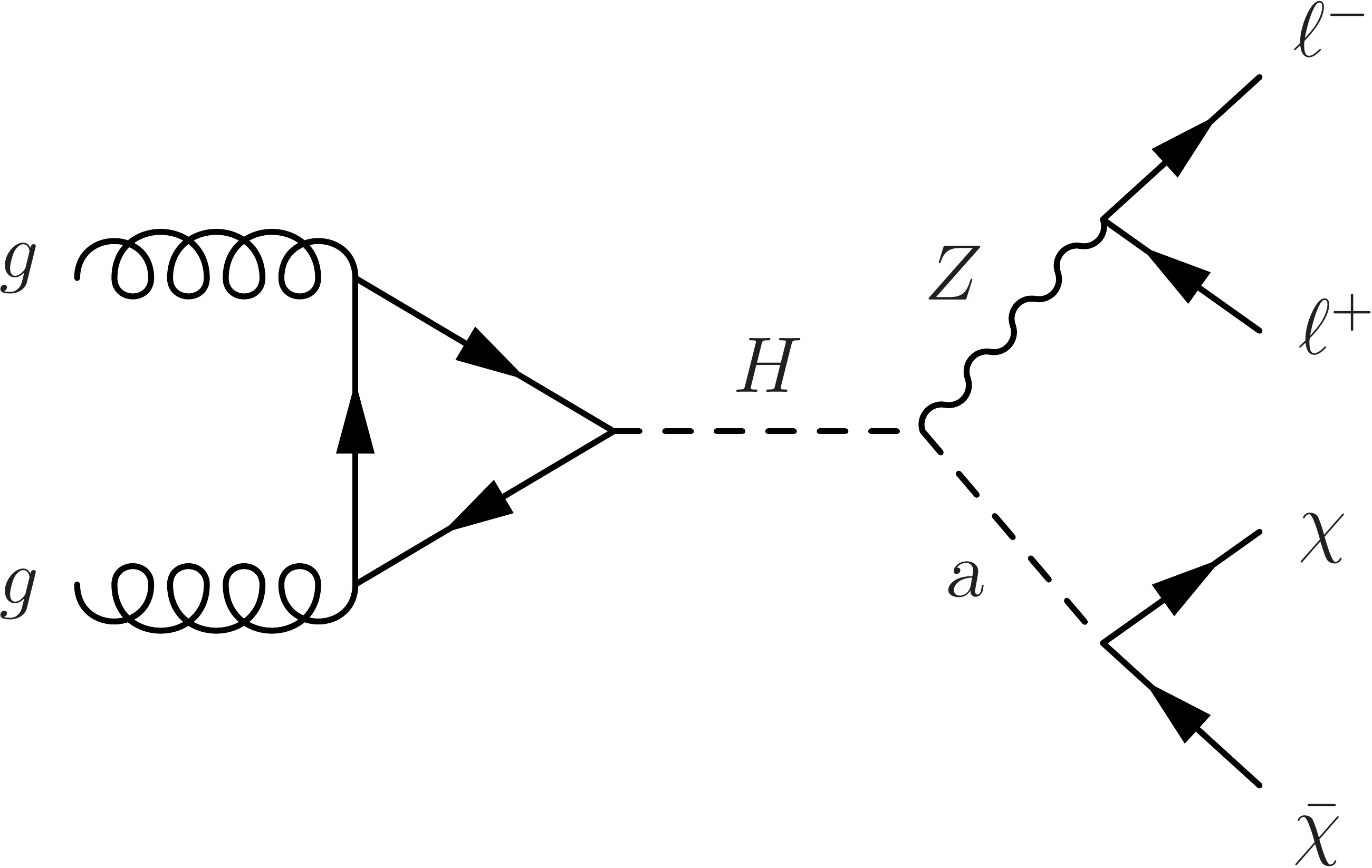}} \\
    {\includegraphics[width=0.38\textwidth]{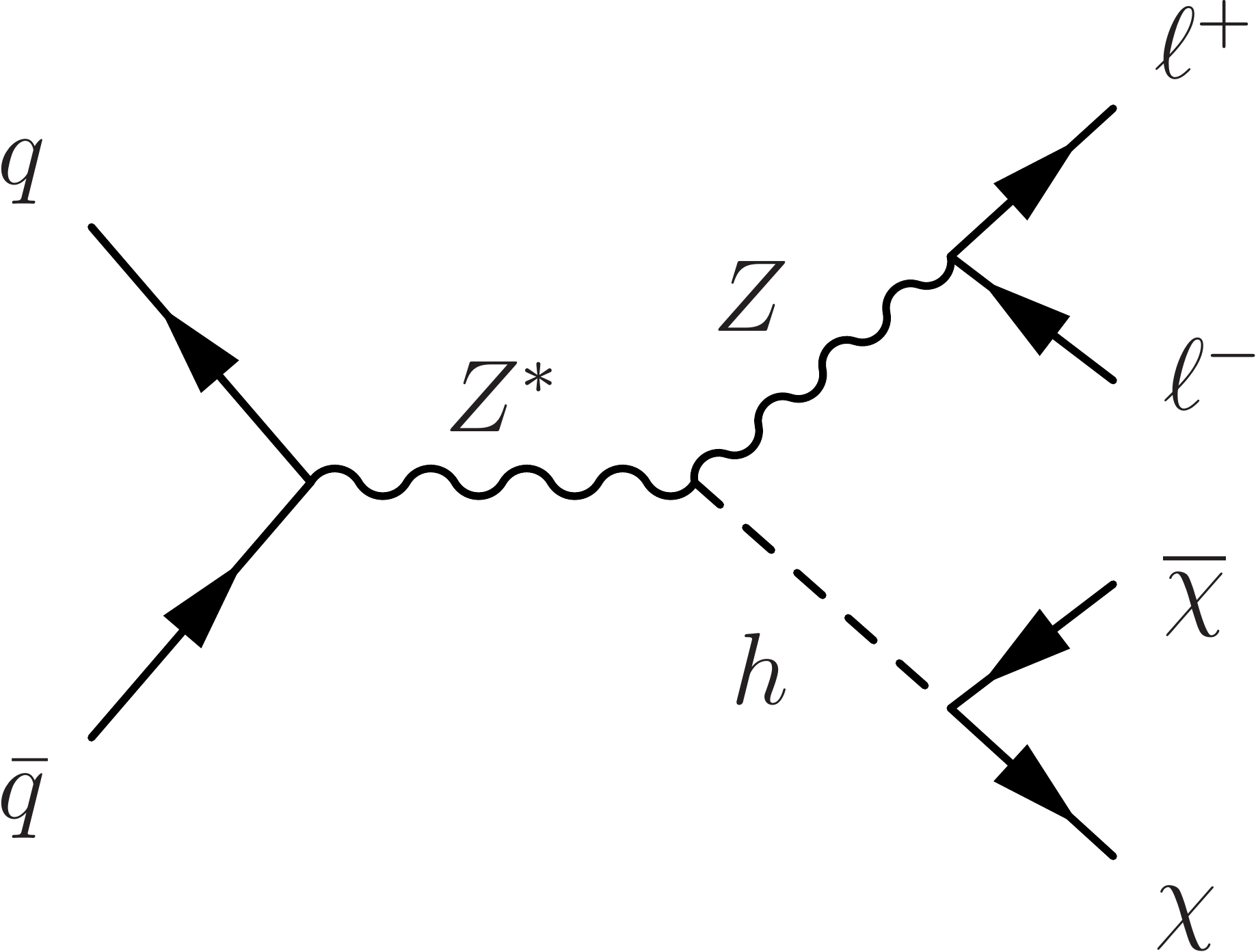}}
    {\includegraphics[width=0.44\textwidth]{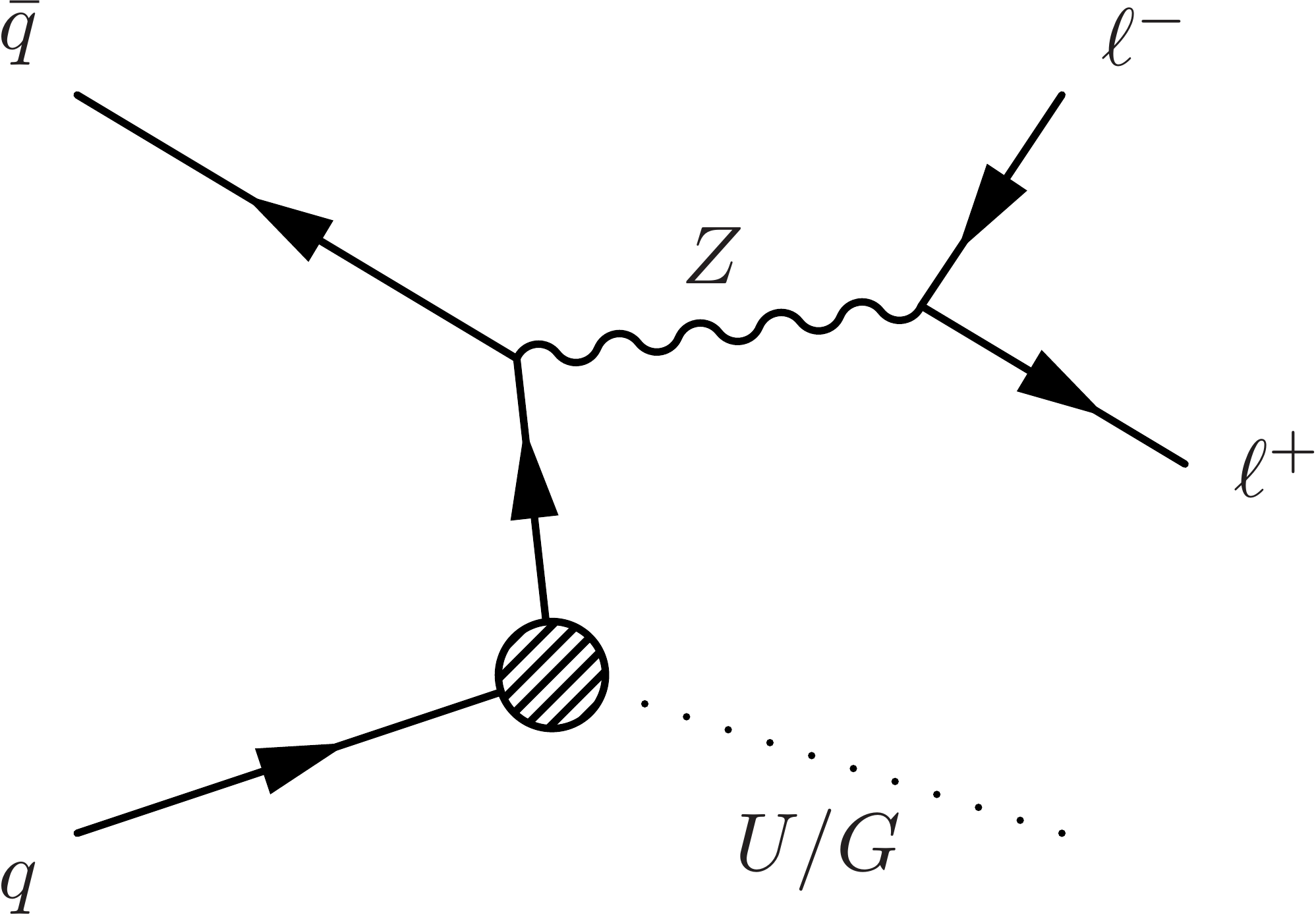}}
  \caption{Feynman diagrams illustrative of the BSM processes that produce
  a final state of a \PZ boson that decays into a pair of leptons and missing transverse momentum:
    (upper left)~simplified dark matter model for a spin-1 mediator,
    (upper right)~2HDM+\aboson model,
    (lower left)~invisible Higgs boson decays, and 
    (lower right)~graviton (\cPG) production in a model with large extra dimensions or unparticle (\textsf{U}) production.
  Here \PSA represents the DM mediator, \PGc represents a DM particle, while (\PH, \Ph) and \aboson represent the 
  scalar and pseudoscalar Higgs bosons, respectively. Here \Ph is identified with the 125\GeV scalar boson.  
  The dotted line represents either an unparticle or a graviton. }
  \label{fig:Feynman}
\end{figure*}

First, we consider a set of simplified models for DM production~\cite{Abercrombie:2015wmb,Boveia:2016mrp}. These models describe the phenomenology of DM production at the LHC
with a small number of parameters and provide a standard for comparing and combining results from different  search 
channels.  Each model  contains a massive mediator exchanged in the $s$-channel, where the mediator (either a vector, 
axial-vector, scalar, or pseudoscalar particle) couples directly to quarks and to the DM particle \PGc. An example tree-level diagram is 
shown in Fig.~\ref{fig:Feynman} (upper left).  The free parameters of each model are
the mass of the DM particle $m_\PGc$,
the mass of the mediator $m_{\text{med}}$, the mediator-quark coupling $g_{\Pq}$, and the mediator-DM coupling 
$g_{\PGc}$. Following the suggestions in Ref.~\cite{Boveia:2016mrp}, for the vector and axial-vector studies, we fix 
the couplings to values of $g_{\Pq}=0.25$ and $g_{\PGc}=1$ and vary the values of $m_\PGc$ and $m_{\text{med}}$, and for the 
scalar and pseudoscalar studies, we fix the couplings $g_{\Pq}=1$ and $g_{\PGc}=1$, set the dark matter particle mass to $m_\PGc=1\GeV$, and vary the values of $m_{\text{med}}$. 
The comparison with data is carried out separately for each of the four spin-parity choices for the mediator.

We also explore a two-Higgs-doublet model (2HDM) with an additional pseudoscalar boson, \aboson, that serves as the mediator 
between DM and ordinary matter.  This ``2HDM+\aboson'' model~\cite{Bauer:2017ota,Abe:2018bpo} 
is a gauge-invariant and renormalizable model that contains a Higgs scalar (\Ph), which we take to be the observed 125 GeV Higgs boson, a heavy
neutral Higgs scalar (\PH), a 
charged Higgs scalar (\Hpm), and two pseudoscalars (\PSA, \aboson), where the pseudoscalar bosons couple to the DM particles. For the process 
studied in this paper, the \PH boson is produced via gluon fusion and decays into a standard model (SM) \PZ boson and the pseudoscalar \aboson. These 
subsequently decay into a pair of leptons and a pair of DM particles, respectively, as shown 
in Fig.~\ref{fig:Feynman} (upper right). The sizable 
couplings of the \PZ boson to the Higgs bosons makes the mono-\PZ channel more sensitive to this model 
than the mono-jet or mono-photon
channels. Among the parameters of this model are the Higgs boson masses,  
the ratio  $\tan\beta$ of the vacuum expectation values of the two Higgs doublets, and  
the mixing angle $\theta$ of the pseudoscalars.
 We consider  only configurations in which  $m_{\PH}=m_{\Hpm}=m_{\PSA}$, 
 and fix the values $\tan\beta=1$ and $\sin\theta=0.35$, following the recommendations of Ref.~\cite{Abe:2018bpo}.

We also examine the case where the \Ph boson acts as a mediator for DM production, as discussed
in ``Higgs portal'' models~\cite{Baek:2012se,Djouadi:2011aa,Djouadi:2012zc,Arcadi:2019lka}.  If $m_{\PGc}<m_{\Ph}/2$,  the Higgs boson 
could decay invisibly into a pair of DM particles. 
 The mechanism for such decays can be found, for example, in many supersymmetric theoretical models that contain a stable 
 neutral lightest supersymmetric particle, \eg, a neutralino~\cite{Belanger:2001am}, that is sufficiently light.
An illustrative Feynman diagram for such a case is shown in Fig.~\ref{fig:Feynman} (lower left), while additional gluon-induced diagrams are also considered.  

In addition to the DM paradigm, we consider a model where unparticles are responsible for the missing transverse momentum in the final state. 
The unparticle physics concept~\cite{Georgi:2007ek,Georgi:2007si} is based on scale
invariance, which is anticipated in many BSM physics 
scenarios~\cite{Kang:2014cia,Rinaldi:2014gha,Cheng:1988zx}.
The effects of the scale-invariant sector (``unparticles'') appear as a non-integral number of invisible massless particles. In this 
scenario, the SM is extended by introducing a scale-invariant Banks--Zaks field, which has a nontrivial 
infrared fixed point~\cite{Banks:1981nn}. 
This field can interact with the SM particles by exchanging heavy particles with a high mass scale 
$M_\textsf{U}$~\cite{Cheung:2007zza}. Below this mass scale, where the coupling is nonrenormalizable, the interaction 
is suppressed by powers of $M_\textsf{U}$ and can be treated within an effective field theory (EFT). The parameters that characterize the unparticle model are 
the possible noninteger scaling dimension of the unparticle operator $d_\textsf{U}$, 
the coupling of the unparticles to SM fields $\lambda$, and the cutoff scale of the EFT $\Lambda_\textsf{U}$. 
In order to remain in the EFT regime, the cutoff scale is set to $\Lambda_\textsf{U}=15\TeV$
and to maintain unitarity, only $d_\textsf{U}>1$ is considered. Figure~\ref{fig:Feynman} (lower right)
shows the tree-level diagram considered in this paper for the production of unparticles associated with a \PZ  boson.

The final SM extension considered in this paper is the Arkani-Hamed--Dimopoulos--Dvali (ADD) model of large extra dimensions~\cite{ArkaniHamed:1998rs,han99:hlz}, 
which is motivated by the disparity between the electroweak (EW) unification 
scale ($M_\text{EW} \sim 100\GeV$) and the Planck scale ($M_\text{Pl} \sim 10^{19}\GeV$).
This model predicts graviton (\cPG) production via the process $\PQq\PAQq \to \PZ + \cPG$, 
as shown in Fig.~\ref{fig:Feynman} (lower right).
The graviton escapes detection, leading to a mono-\PZ signature. In the ADD model, the apparent Planck scale
 in four spacetime dimensions is given by $M_\text{Pl}^2 \approx M_{\mathrm{D}}^{n+2}R^n$, 
 where $M_{\mathrm{D}}$ is the fundamental Planck scale in
the full ($n$+4)-dimensional spacetime and $R$ is the compactification length scale of the extra
dimensions. Assuming $M_{\mathrm{D}}$ is of the same order as $M_\text{EW}$, the observed large value
of $M_\text{Pl}$ suggests values of $R$ much larger than the Planck length. These values are on the order of nm for $n=3$, 
decreasing with larger values of $n$. The consequence of the large compactification scale is that 
the mass spectrum of the Kaluza--Klein graviton states becomes nearly continuous\cite{ArkaniHamed:1998rs,han99:hlz}, resulting in a broadened spectrum for the transverse momentum (\pt) of the \PZ boson.

\section{The CMS detector}\label{sec:CMS_Detector}

The central feature of the CMS apparatus is a superconducting solenoid of 6\unit{m} internal diameter, providing a magnetic 
field of 3.8\unit{T}. Within the solenoid volume are a silicon pixel and strip tracker, a lead tungstate crystal electromagnetic 
calorimeter (ECAL), and a brass and scintillator hadron calorimeter (HCAL), each composed of a barrel and two endcap 
sections. Forward calorimeters extend the pseudorapidity ($\eta$) coverage provided by the barrel and endcap detectors. Muons are 
detected in gas-ionization chambers embedded in the steel flux-return yoke outside the solenoid.

Events of interest are selected using a two-tiered trigger system~\cite{Khachatryan:2016bia}. The first level (L1), composed of 
custom hardware processors, uses information from the calorimeters and muon detectors to select events at a rate of around 
100\unit{kHz} within a time interval of less than 4\mus. The second level, known as the high-level trigger (HLT), consists of a 
farm of processors running a version of the full event reconstruction software optimized for fast processing, and reduces the 
event rate to around 1\unit{kHz} before data storage. 

A more detailed description of the CMS detector, together with a definition of the coordinate system used and the relevant 
kinematic variables, can be found in Ref.~\cite{Chatrchyan:2008zzk}. 

\section {Data samples and event reconstruction}
\label{sec:objects}

This search uses $\Pp\Pp$ collision events collected with the CMS 
detector during 2016, 2017, and 2018 corresponding to a total integrated 
luminosity of \usedLumi.
The data sets from the three different years are analyzed independently with appropriate calibrations 
and corrections to take into account the different LHC running conditions and 
CMS detector performance.

Several SM processes can contribute to the mono-\PZ signature. 
The most important backgrounds come from diboson processes:  
$\WZ\to\ell\PGn\ell\ell$ where one lepton escapes detection, $\ZZ\to\ell\ell\PGn\PGn$, 
and $\WW\to\ell\ell\PGn\PGn$.  There can also be contributions 
where energetic leptons are produced by decays of top quarks in 
$\ttbar$ or $\PQt\PW$ events.  Smaller contributions may come from 
triple vector boson processes ($\PW\PW\PZ$, $\PW\PZ\PZ$, and $\PZ\PZ\PZ$), 
$\ttbar\PW\to\PW\PW\bbbar\PW$, $\ttbar\PZ\to\PW\PW\bbbar\PZ$, and $\ttbar\PGg\to\PW\PW\bbbar\PGg$, 
 referred to collectively as \PV\PV\PV due to the similar decay products. Drell--Yan  (DY) production of lepton pairs, 
$\PZ/\PGg^*\to\ell\ell$, has no intrinsic source of missing transverse momentum but can still mimic a mono-\PZ signature
when the momentum of the recoiling system is poorly measured. 
A minor source of background is from events with a vector boson and a misreconstructed photon, referred to as $\PV\PGg$.

Monte Carlo simulated events are used to model the expected signal and background 
yields. Three sets of simulated events for each process are 
used in order to match the different data taking conditions.
The samples for DM production are generated using
 the \DMSIMP package~\cite{Mattelaer_2015,Neubert_2016}
interfaced with $\MGvATNLO$ 2.4.2 \cite{Alwall:2014hca,Frederix2012,Artoisenet:2012st,Alwall:2007fs}.
The pseudoscalar and scalar model samples are generated at leading order (LO) in quantum chromodynamics (QCD), while the 
vector and axial-vector model samples are generated at next-to-leading-order (NLO) in QCD.
The $\POWHEG$v2~\cite{Nason:2004rx,Frixione:2007vw,Alioli:2010xd,Frixione:2007nw,Bagnaschi:2011tu}
generator is used to simulate the $\PZ\Ph$ signal process of the invisible Higgs boson at NLO in QCD, as well as the $\ttbar$, $\PQt\PW$, 
and diboson processes. The BSM Higgs boson production cross sections, as a function of the 
Higgs boson mass for the $\PZ\Ph$ process are taken from Refs.~\cite{deFlorian:2016spz}. 
Samples for the 2HDM+\aboson model are generated at NLO with $\MGvATNLO$ 2.6.0. 
Events for both the ADD and unparticle models are generated at LO using an EFT 
implementation in $\PYTHIA$ 8.205 in 2016 and 8.230 in 2017 and 2018~\cite{Sjostrand:2014zea,Ask:2009pv}. 
In order to ensure the validity of the effective theory used in the ADD model, a truncation method, 
described in Ref.~\cite{Ask_2009}, is applied. Perturbative calculations are only valid in cases where the square of the
 center-of-mass energy ($\hat{s}$) of the incoming partons is smaller than the fundamental scale of the theory 
 ($M_{\mathrm{D}}^2$). As such, this truncation method suppresses the cross section 
for events with $\hat{s} > M_\text{D}^2$ by a factor of $M_\text{D}^4/\hat{s}^{2}$. 
The effect of this truncation is largest for small values of $M_{\mathrm{D}}$, but also increases with 
the number of dimensions $n$ as more energy is lost in extra dimensions. 
The $\MGvATNLO$ 2.2.2 (2.4.2) generator in 2016 (2017 and 2018) is used for the simulation 
of the \PV\PV\PV, $\PV\PGg$, and DY samples, at NLO accuracy in QCD.

{\tolerance=800
The set of parton distribution functions (PDFs) used for simulating the 2016 sample is NNPDF 3.0 NLO~\cite{Ball:2017nwa} 
and for the 2017 and 2018 samples it is NNPDF 3.1 NNLO. 
For all processes, the parton showering and hadronization are simulated using 
\PYTHIA 8.226 in 2016 and 8.230 in 2017 and 2018.
The modeling of the underlying event is generated using the 
CUETP8M1~\cite{Khachatryan:2015pea} (CP5~\cite{Sirunyan:2019dfx})
for simulated samples corresponding to the 2016 (2017 and 2018) data sets. The only exceptions to this are the 
2016 top quark sample, which uses CUETP8M2~\cite{Khachatryan:2015pea} and the simplified DM (2HDM+\aboson) samples, 
which uses CP3~\cite{Sirunyan:2019dfx} (CP5) tunes for all years.
All events are processed through a simulation of the CMS detector based on 
\GEANTfour~\cite{Agostinelli:2002hh} and are reconstructed with the same algorithms as 
used for data. Simultaneous $\Pp \Pp$ collisions in the same or nearby bunch 
crossings, referred to as pileup, are also simulated. 
The distribution of the number of such interactions in the simulation is chosen to match the data, with periodic adjustments to take account of changes in LHC operating conditions~\cite{Sirunyan:2020foa}. 
The average number of pileup interactions was 23 for the 2016 data and 32 for the 2017 and 2018 data.
\par}

Information from all subdetectors is combined and used by the CMS particle-flow (PF) 
algorithm~\cite{Sirunyan:2017ulk} for particle reconstruction and identification.
The PF algorithm aims to reconstruct and identify each individual particle in an event, 
with an optimized combination of information from the various elements of the CMS detector. 
The energies of photons are obtained from the ECAL measurement. The energies of electrons 
are determined from a combination of the electron momentum at the primary interaction 
vertex as determined by the tracker, the energy from the corresponding ECAL cluster, 
and the energy sum from all bremsstrahlung photons spatially compatible with originating 
from the electron track. The momentum of muons is obtained from the curvature of the corresponding 
track in the tracker detector in combination with information from the muon stations. 
The energies of charged hadrons are determined from a combination 
of their momentum measured in the tracker and the matching ECAL and HCAL energy deposits, 
corrected for the response function of the calorimeters to hadronic showers. Finally, 
the energies of neutral hadrons are obtained from the corresponding corrected ECAL and HCAL energies.

The candidate vertex with the largest value of summed physics-object $\pt^2$ is taken 
to be the primary $\Pp\Pp$ interaction vertex. The physics objects are the jets, 
clustered using the jet finding algorithm~\cite{Cacciari:2008gp,Cacciari:2011ma} with 
the tracks assigned to candidate vertices as inputs, and the associated missing 
transverse momentum, taken as the negative vector sum of the \pt of those jets.

Both electron and muon candidates must pass certain identification criteria to be further 
selected in the analysis. They must satisfy requirements on the transverse momentum
and pseudorapidity: $\pt > 10\GeV$ and 
$\abs{\eta} < 2.5~(2.4)$ for electrons (muons). At the final level, 
a medium working point~\cite{Khachatryan:2015iwa,Sirunyan:2018fpa} 
is chosen for the identification criteria, including requirements on the impact parameter 
of the candidates with respect to the primary vertex and their isolation with 
respect to other particles in the event. The efficiencies for these selections are 
about 85 and 90\% for each electron and muon, respectively.

In the signal models considered in this paper, the amount of hadronic activity tends to be small, so events with multiple clustered jets are vetoed.
For each event, hadronic jets are clustered from reconstructed particle candidates using the infrared and collinear safe 
anti-\kt algorithm~\cite{Cacciari:2008gp, Cacciari:2011ma} with a distance parameter of 0.4. 
Jet momentum is determined as the vectorial sum of all particle momenta in the jet, 
and is found from simulation to be, on average, within 5 to 10\% of the true momentum 
over the entire spectrum and detector acceptance. Pileup interactions  
can contribute additional 
tracks and calorimetric energy depositions to the jet momentum. To mitigate this effect, 
charged particles identified to be originating from pileup vertices are discarded and an 
offset is applied to correct for remaining contributions~\cite{Khachatryan:2016kdb}. 
Jet energy corrections are derived from simulation to bring the measured response of jets to the average
of simulated jets clustered from the generated final-state particles. 
In situ measurements of the momentum balance in dijet, 
photon+jet, $\PZ$+jet, and multijet events are used to 
determine corrections for residual differences between jet energy scale in data and 
simulation~\cite{Khachatryan:2016kdb}. The jet energy resolution amounts typically to 
15\% at 10\GeV, 8\% at 100\GeV, and 4\% at 1\TeV. Additional selection criteria are 
applied to each jet to remove jets potentially dominated by anomalous contributions from some subdetector components or reconstruction failures~\cite{CMS:2017wyc}. 
Jets with $\pt> 30\GeV$ and $\abs{\eta}<4.7$ are considered for the analysis. 

To identify jets that originated from \cPqb quarks, we use the medium working point of the 
DeepCSV algorithm~\cite{Sirunyan:2017ezt}. This selection was chosen to remove events from top quark decays originating specifically from $\ttbar$ production, without causing a significant loss of signal. For this working point, 
the efficiency to select b~quark jets is about 70\% and the probability for mistagging jets originating from the hadronization of gluons or $\cPqu/\cPqd/\cPqs$ quarks is about 1\% in simulated $\ttbar$ events.

To identify hadronically decaying \PGt leptons (\tauh), we use the hadron-plus-strips 
algorithm~\cite{Sirunyan:2018pgf}. This algorithm
constructs candidates seeded by PF jets that are consistent with either
a single or triple charged pion decay of the \PGt lepton. In the
single charged pion decay mode, the presence of neutral pions is
detected by reconstructing their photonic decays. Mistagged jets originating from non-\PGt decays are rejected by
a discriminator that takes into account the pileup contribution to the
neutral component of the \tauh decay~\cite{Sirunyan:2018pgf}. 
The efficiency to select real hadronically decaying \PGt leptons is about 75\% and the 
probability for mistagging jets is about 1\%.

The missing transverse momentum vector \ptvecmiss is computed as the negative vector sum of the transverse momenta of 
all the PF candidates in an event, and its magnitude is denoted as \ptmiss~\cite{Sirunyan:2019kia}. The \ptvecmiss is 
modified to account for corrections to the energy scale of the reconstructed jets in the event. 
Events with anomalously high \ptmiss can originate from a variety of reconstruction failures, detector malfunctions, or 
noncollision backgrounds. Such events are
rejected by event filters that are designed to identify more than 85--90\% of the spurious high-\ptmiss events with 
a misidentification rate of less than 0.1\%~\cite{Sirunyan:2019kia}. 
\section{Event selection}
\label{sec:selection}

Events with electrons (muons) are collected using
dielectron (dimuon) triggers, with thresholds of
$\pt > 23$ (17)\GeV and $\pt > 12$ (8)\GeV for the electron (muon) with the highest and second-highest measured \pt, respectively.
Single-electron and single-muon triggers with \pt thresholds of 25 (27) and 20 (24)\GeV for 2016 (2017--2018) are used to recover residual inefficiencies, ensuring a trigger efficiency above 99\% for events passing the offline selection.

In the signal region (SR), events are required to have two ($N_{\ell} = 2$) well-identified, isolated electrons or muons with
the same flavor and opposite charge ($\Pep\Pem$ or $\PGmp\PGmm$). 
At least one electron or muon of the pair must have $\pt > 25\GeV$, 
while the second must have $\pt > 20\GeV$. 
In order to reduce nonresonant background, the dilepton invariant mass is required to be within 
15\GeV of the world-average \PZ boson mass $m_{\PZ}$~\cite{Tanabashi:2018oca}. 
Additionally, we require the \pt of the dilepton system $\pt^{\ell\ell}$  
to be larger than 60\GeV to reject the bulk
of the DY background. 
Since little hadronic activity is expected for the signal, we reject events 
having more than one jet with $\pt>30\GeV$ within $\abs{\eta}<4.7$. 
The top quark background is further suppressed by 
rejecting events containing any \PQb-tagged jet with $\pt > 30\GeV$
reconstructed within the tracker acceptance of $\abs{\eta} < 2.4$.
To reduce the \WZ background in which both bosons decay
leptonically, we remove events containing additional electrons or muons with loose identification and  
with $\pt > 10\GeV$. Events containing a loosely identified \tauh candidate with $\pt>18\GeV$ and $\abs{\eta} < 2.3$ are also rejected. Decays that are consistent with production of muons or electrons are rejected by an overlap veto.

In addition to the above criteria, there are several selections designed to further reduce the SM background.
The main discriminating variables are: the missing transverse momentum, \ptmiss;
the azimuthal angle formed between the dilepton \pt and
the \ptvecmiss, $\Delta\phi(\ptvecll,\ptvecmiss)$; and the
balance ratio, $\abs{\ptmiss-\pt^{\ell\ell}}/\pt^{\ell\ell}$. The latter two variables are especially powerful in rejecting DY and top quark 
processes. Selection criteria are optimized to obtain the best signal sensitivity for the range of DM processes considered.
The final selection requirements are: 
$\ptmiss > 100\GeV$, $\Delta\phi(\ptvecll,\ptvecmiss) > 2.6\unit{radians}$, and 
$\abs{\ptmiss-\pt^{\ell\ell}}/\pt^{\ell\ell} < 0.4$.  

{\tolerance=9600
For the 2HDM+\aboson model, the selection differs slightly.  We make a less stringent requirement on the 
missing transverse momentum, $\ptmiss>80\GeV$, and require the transverse mass, 
$\mT=\sqrt{\smash[b]{2p^{\mathrm{T}}_{\ell\ell}\ptmiss[1-\cos\Delta\phi(\ptvecll,\ptvecmiss)]}}$
to be greater than 200\GeV.
The kinematic properties of the 2HDM+\aboson production yield a peak in the \mT spectrum near the neutral Higgs scalar (H) mass 
that is advantageous for background discrimination.
\par}

In order to avoid biases in the \ptmiss calculation due to jet mismeasurement, 
events with one jet are required to have the azimuthal angle between this jet and the missing transverse  momentum, $\Delta\phi(\ptvec^{\mathrm{j}},\ptvecmiss)$, larger than 0.5\unit{radians}.
To reduce the contribution from backgrounds such as \WW and \ttbar, we apply 
a requirement on the distance between the
two leptons in the $(\eta,\phi)$ plane, $\Delta R_{\ell\ell} < 1.8$, where $\Delta R_{\ell\ell} = \sqrt{\smash[b]{(\Delta\phi_{\ell\ell})^2+(\Delta\eta_{\ell\ell})^2}}$.

A summary of the selection criteria for the SR is given in Table~\ref{tab:selectioncuts}.  

\begin{table*}[hbtp]
  \centering
  \topcaption{Summary of the kinematic selections for the signal region. \label{tab:selectioncuts}}
 {
  \begin{tabular} {lcc}
\hline
Quantity & Requirement  & Target backgrounds \\
\hline
$N_{\ell}$                                                     & ${=}2$ with additional lepton veto                                          & \WZ, $\PV\PV\PV$      \\
$\pt^{\ell}$                                                   & ${>}$25/20\GeV for leading/subleading             & Multijet\\
Dilepton mass                                   & $\left|m_{\ell\ell}-m_{\PZ}\right| < 15\GeV$    & \WW, top quark         \\
Number of jets                                                 & ${\leq}1$ jet with $\pt^{\mathrm{j}} > 30\GeV$    & DY, top quark, $\PV\PV\PV$ \\
$\pt^{\ell\ell}$                                               & ${>}60\GeV$                                    & DY          \\
\PQb tagging veto                                                 & 0 \PQb-tagged jet with $\pt>30\GeV$                           & Top quark, $\PV\PV\PV$   \\
$\PGt$ lepton veto                                             & 0 \tauh\ cand. with $\pt^{\PGt}>18\GeV$ & \WZ   \\
$\Delta \phi(\ptvec^{\mathrm{j}},\ptvecmiss)$                  & ${>}0.5\unit{radians}$                                  & DY, \WZ         \\
$\Delta \phi(\ptvecll,\ptvecmiss)$           & ${>}2.6\unit{radians}$ 	                              & DY          \\
$\abs{\ptmiss-\pt^{\ell\ell}}/\pt^{\ell\ell}$                         & ${<}0.4$ 	                              & DY             \\
$\Delta R_{\ell\ell}$                                          & ${<}1.8$ 	                              & \WW, top quark       \\
\ptmiss (all but 2HDM+\aboson)                           & ${>}100\GeV$  				      & DY, \WW, top quark   \\
\ptmiss (2HDM+\aboson only)                           & ${>}80\GeV$  				      & DY, \WW, top quark   \\
\mT (2HDM+\aboson only)                           & ${>}200\GeV$  				      & DY, \WW, \ZZ, top quark   \\
  \hline
  \end{tabular}
}
\end{table*}

\section{Background estimation}
\label{sec:backgrounds}
We estimate the background contributions using combined information from simulation and control regions (CRs) in data.
 A simultaneous maximum likelihood fit to the \ptmiss or \mT distributions in the SR and CRs constrains the background normalizations and their uncertainties. 
Specific CRs target different categories of background processes, as described below.

\subsection{The three-lepton control region}
The ${\WZ\to\ell'\PGn\ell\ell}$ decay mode can contribute to the SR when the third lepton ($\ell'=\Pe$ or $\PGm$) 
escapes detection, and this same process can be monitored in an orthogonal CR, where the third lepton is 
identified and then removed.
The construction of the three-lepton ($3\ell$) CR is based on events with three well-reconstructed charged leptons.
A \PZ boson candidate is selected in the same manner as for the SR , while an additional electron or muon with identical quality and isolation is required. 
In cases where there are multiple \PZ boson candidates, the candidate with invariant mass closest to the \PZ boson mass is selected.
To enhance the purity of the \WZ selection, \ptmiss of at least 30\GeV is required and the invariant mass of three leptons is required to be larger than 100\GeV. 
The backgrounds in this CR are similar to those in the SR, with a sizable nonprompt background from DY events where a jet is misidentified as a lepton~\cite{Sirunyan:2019bez}.
An additional minor source of background is from events with a vector boson and a misreconstructed photon ($\PV\PGg$). All background estimates for this CR are taken from simulation.

\begin{figure}[htbp]
\centering
\includegraphics[width=0.48\textwidth]{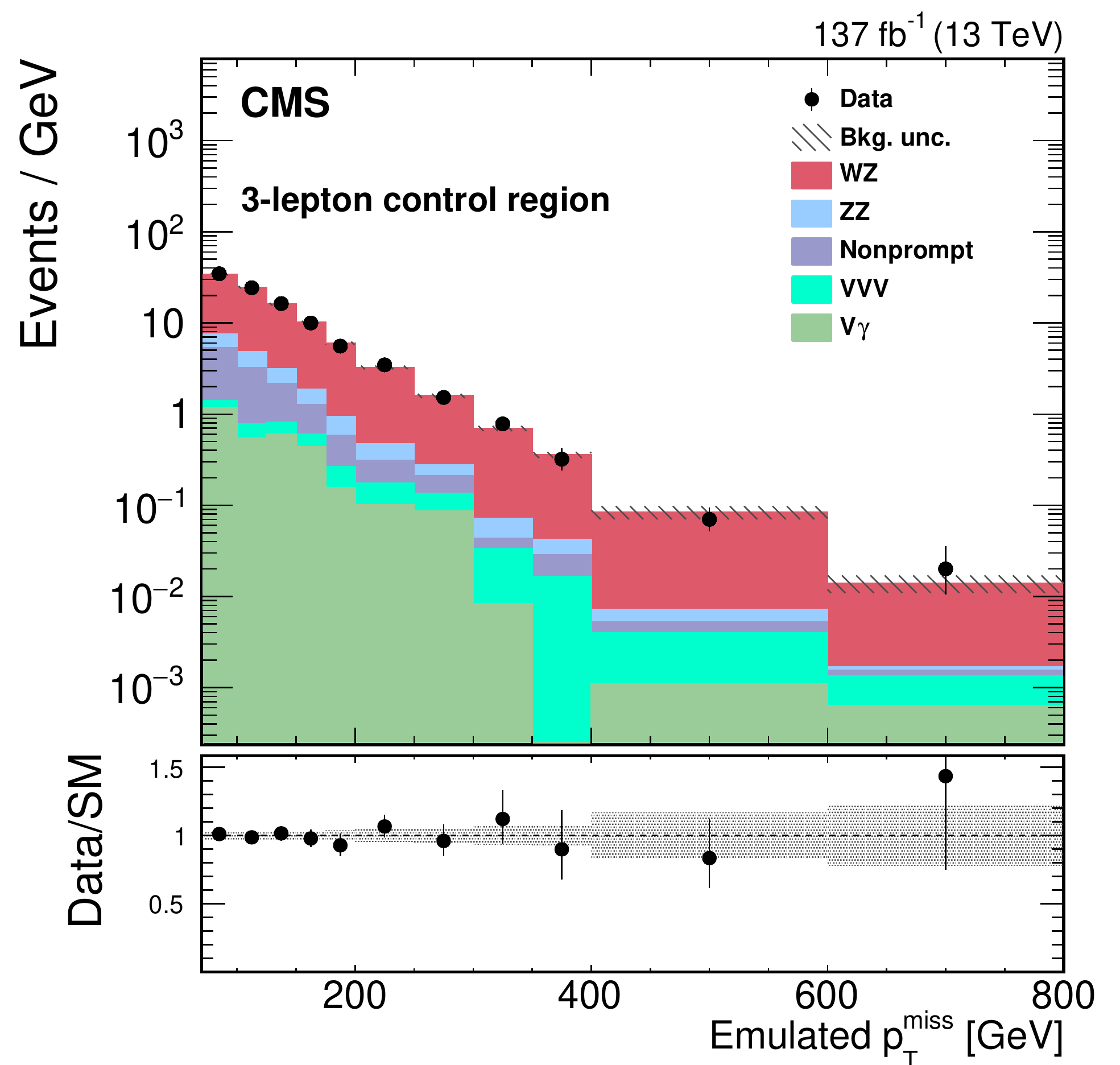}
\includegraphics[width=0.48\textwidth]{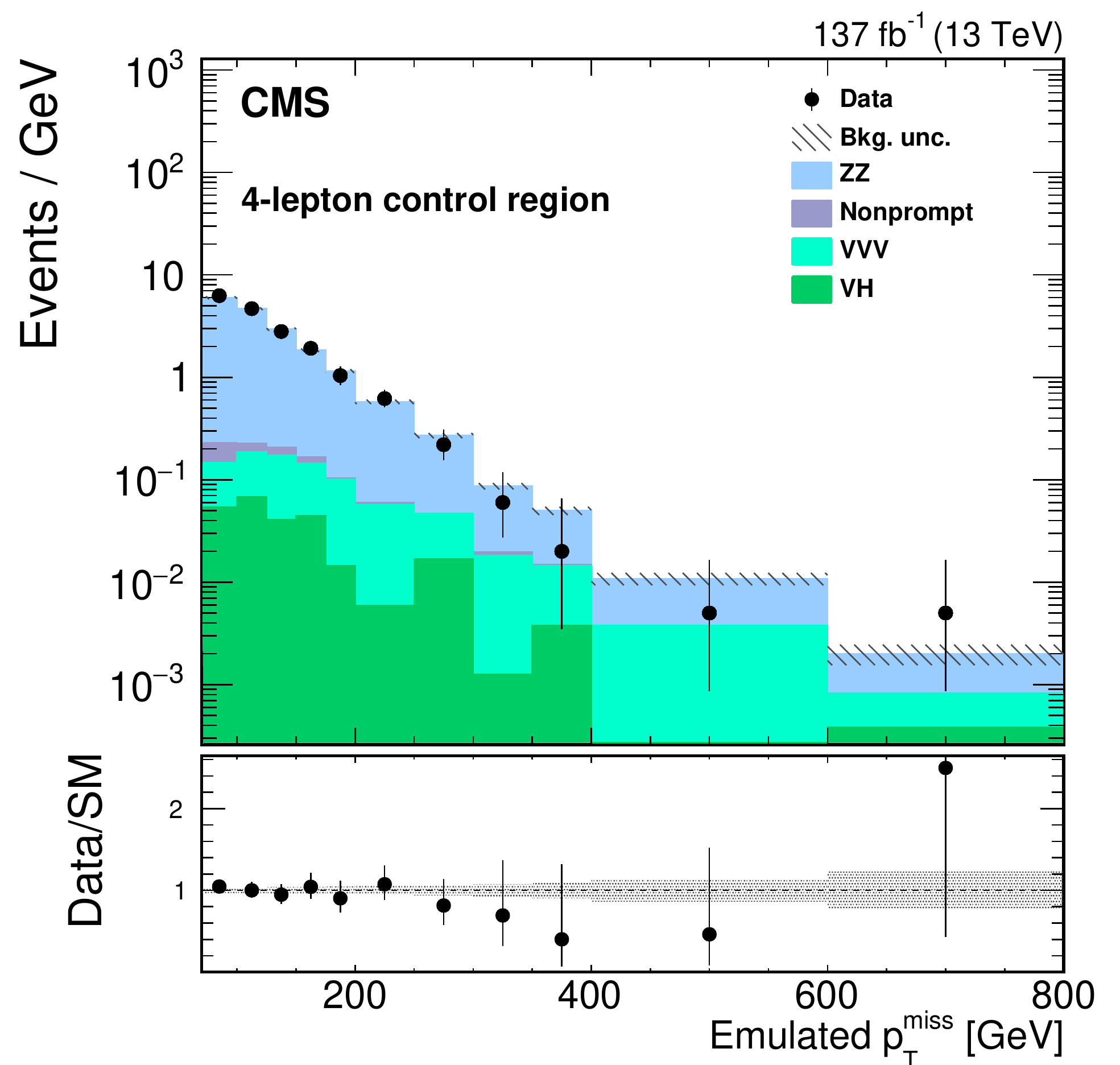}
\caption{
  Emulated \ptmiss distribution in data and simulation for the $3\ell$ (\cmsLeft) and $4\ell$ (\cmsRight) 
  CRs. Uncertainty bands correspond to the postfit combined statistical and systematic components, where the fitting method is described in Section~\ref{sec:fitting}.
} 
\label{fig:histo_fakemet}
\end{figure}

To simulate the consequences of not detecting the third lepton, the ``emulated \ptmiss'' is estimated from the vectorial sum of  \ptvecmiss
and the transverse momentum (\ptvec) of the additional lepton.
The emulated \ptmiss is then used in place of the reconstructed \ptmiss and the same selection is applied as for the SR.
Since there is negligible contamination from $\WZ\to\PGt\PGn\ell\ell$ and top quark backgrounds in this CR,
no veto is applied on additional \tauh or \PQb jet candidates. The resulting emulated \ptmiss spectrum is shown in Fig.~\ref{fig:histo_fakemet} (\cmsLeft). 
For the 2HDM+\aboson case, the ``emulated \mT'' is used instead of ``emulated \ptmiss'' with the same selections. 

\subsection{The four-lepton control region}
The \ZZ process contributes to the SR through the
$\ZZ\to\ell\ell\PGn\PGn$ decay mode, and the same production process can be monitored via the decay 
mode $\ZZ\to 4\ell$.
The $4\ell$ CR is based on events with two pairs of charged leptons.  Each pair comprises two leptons of opposite charge and the same flavor and corresponds to a \PZ candidate.  
Two of the four leptons must fulfill the same requirements on the leptons as in the SR, while, in order to increase the yield,  the other two leptons need only pass relaxed lepton quality requirements.
The highest \pt \PZ boson candidate is required to have an invariant mass within 
35\GeV of the \PZ boson mass $m_{\PZ}$~\cite{Tanabashi:2018oca}. 
Additionally, we require the transverse momentum of this \PZ boson candidate  
to be larger than 60\GeV. 
Additional backgrounds to the \ZZ final state are events from triboson processes, events with a vector boson and a higgs boson ($\PV\Ph$) and from nonprompt events. 
These backgrounds are almost negligible. All background estimates for this CR are taken from simulation.

For these four-lepton events, the emulated \ptmiss is calculated as the vectorial sum of the $\ptvecmiss$ and the \ptvec of the 
\PZ boson candidate with the larger absolute mass difference to $m_{\PZ}$.
The choice of which \PZ boson to use as a proxy for an invisibly decaying boson negligibly alters the emulated \ptmiss spectrum.
The same selection as the SR is then applied using the emulated \ptmiss in place of the reconstructed \ptmiss, with the exception of the \tauh and \PQb jet candidate vetoes.
The resulting emulated \ptmiss spectrum is shown in Fig.~\ref{fig:histo_fakemet} (\cmsRight).
Similarly to the $3\ell$ CR, the ``emulated \mT'' is used instead of ``emulated \ptmiss'' for the 2HDM+\aboson case and the distribution is well described by the SM background estimations.

\subsection{The electron-muon control region}
We estimate the contribution of the flavor-symmetric backgrounds from an $\Pe\PGm$ CR based on events with two leptons of different flavor and opposite charge 
($\Pe^{\pm}\PGm^{\mp}$) that pass all other analysis selections.
This CR is largely populated by nonresonant backgrounds (NRB) consisting mainly of leptonic \PW\ boson decays in
$\ttbar$, $\cPqt\PW$, and \WW events, where the dilepton mass happens to fall inside the $\PZ$ boson mass window.
Small contributions from single top quark events produced via
$s$- and $t$-channel processes, and $\PZ\to \PGt\PGt$
events in which \PGt leptons decay into light leptons and neutrinos, are also
considered in the NRB estimation. 

\subsection{The DY control region}
The DY background is dominant in the region of low \ptmiss.
This process does not produce undetectable particles. Therefore, any nonzero \ptmiss arises from
mismeasurement or limitations in the  detector acceptance.
The estimation of this background uses simulated DY events, for which the normalization is taken from data in a sideband CR
of $80 < \ptmiss  < 100\GeV$ where the signal contamination is negligible, with all other selections applied. For the 2HDM+\aboson analysis, a similar approach is taken with relaxed $\ptmiss$ selection of $50 < \ptmiss  < 100\GeV$ and an additional selection of $\mT < 200\GeV$ applied. The sideband CR is included in the maximum likelihood fit and
a 100\% uncertainty is assigned  to the extrapolation from this CR to the SR.
This uncertainty has little effect on the results because of the smallness of the overall contribution
 from the DY process in the SR.

\section{Fitting method}\label{sec:fitting}
After applying the selection, we perform a binned 
maximum likelihood fit to discriminate between the potential signal and the 
remaining background processes. 
The data sets for each data-taking year are kept separate in the fit.
 This yields a better expected significance than combining them into a single set because
the signal-to-background 
ratios are different for the three years due to the different data-taking conditions. The 
electron and muon channels have comparable signal-to-background ratios, and
 are combined in the fit, while the contributions, corrections and systematic uncertainties are calculated individually.

The \ptmiss distribution of events passing the selection is used as the discriminating variable in the fit for all of the 
signal hypotheses except for the 2HDM+\aboson model. 
For this model, the \mT distribution is used since a Jacobian peak around the pseudoscalar Higgs boson mass is expected.
Events in the SR are split into 0-jet and 1-jet categories to take into 
account the different signal-to-background ratios. 
In addition, for the CRs defined in Section~\ref{sec:backgrounds}, events with 0-jet and 1-jet are included as a single category in the fit.
The $\Pe\PGm$ and DY CRs
are each included as a single bin  corresponding to the total yield.  
The \ptmiss or \mT spectra  in the 
$3\ell$ and $4\ell$ CRs are included in the fit with the same binning
 as in the SR, where  these spectra are based upon the emulated \ptmiss.
To allow for further freedom in the \ZZ and \WZ background estimation,  the
\ptmiss and emulated \ptmiss distributions are split into three regions with independent normalization parameters: low
 (${<}200\GeV$), medium (200--400\GeV), 
and high (${>}400\GeV$), with uncertainties of 10, 20, and 30\%, respectively. 
These values are based on the magnitudes of the theoretical uncertainties as described in Section~\ref{sec:systematics}. 
For fits to the 2HDM+\aboson model, three similar \mT 
regions are chosen with the same uncertainties: low (${<}400\GeV$), medium (400--800\GeV),
and high (${>}800\GeV$). To make the best use of the statistical power in the CRs and to 
take advantage of the similarities of the production processes, we take the normalization
factors to be correlated for the \ZZ and the \WZ backgrounds in each \ptmiss region. 

For each individual bin, a Poisson likelihood term describes the fluctuation 
of the data around the expected central value, which is given by the sum of the contributions 
from signal and background processes. Systematic uncertainties are represented by
nuisance parameters $\theta$ with log-normal probability density functions used for normalization uncertainties and Gaussian functions used for shape-based uncertainties, with the functions centered on their 
nominal values $\hat{\theta}$. The uncertainties affect 
the overall normalizations of the signal and background templates, as well as the shapes 
of the predictions across the distributions of observables. Correlations among
systematic uncertainties in different categories are taken into account as discussed in Section~\ref{sec:systematics}. The total 
likelihood is defined as the product of the likelihoods of the individual bins and 
the probability density functions for the nuisance parameters:
\begin{equation}
\mathcal{L} =  \mathcal{L}_{\text{SR}}  \mathcal{L}_{3\ell}  \mathcal{L}_{4\ell} 
\mathcal{L}_{\Pe\PGm}  \mathcal{L}_{\text{DY}} \, f_{\text{NP}} \Big(\theta \mid \hat{\theta} \Big)
\end{equation}
The factors of the likelihood can be written more explicitly as 
\begin{align}
\ifthenelse{\boolean{cms@external}}
{\begin{split}
\mathcal{L} _{\text{SR}}=  & \prod_{i,j} \mathcal{P} \Big( N^{\text{SR}}_{\text{obs},i,j} \mid 
\mu_{\text{DY}}N^{\text{SR}}_{\text{DY},i,j}(\theta) 
\\ & + \mu_{\text{NRB}}N^{\text{SR}}_{\text{NRB},i,j}(\theta) + 
N^{\text{SR}}_{\text{other},i,j}(\theta) 
\\ & + \mu_{\PV\PV,r(i)}(N^{2\ell}_{\ZZ,i,j}(\theta) + N^{\text{SR}}_{\WZ,i,j}(\theta)) 
\\ &+  \mu N^{\text{SR}}_{\text{Sig},i,j}(\theta) \Big),
\end{split}}
{\begin{split}
\mathcal{L} _{\text{SR}}=  & \prod_{i,j} \mathcal{P} \Big( N^{\text{SR}}_{\text{obs},i,j} \mid 
\mu_{\text{DY}}N^{\text{SR}}_{\text{DY},i,j}(\theta) + \mu_{\text{NRB}}N^{\text{SR}}_{\text{NRB},i,j}(\theta) + 
N^{\text{SR}}_{\text{other},i,j}(\theta) 
\\ & + \mu_{\PV\PV,r(i)}(N^{2\ell}_{\ZZ,i,j}(\theta) + N^{\text{SR}}_{\WZ,i,j}(\theta)) + 
\mu N^{\text{SR}}_{\text{Sig},i,j}(\theta) \Big),
\end{split}}
\\
\mathcal{L} _{3\ell}= & \prod_{i} \mathcal{P} \Big( N^{3\ell}_{\text{obs},i} \mid N^{3\ell}_{\text{other},i}(\theta) + 
\mu_{\PV\PV,r(i)} N^{3\ell}_{\WZ,i}(\theta) \Big),
\\
\mathcal{L} _{4\ell}=& \prod_{i} \mathcal{P} \Big( N^{4\ell}_{\text{obs},i} \mid N^{4\ell}_{\text{other},i}(\theta) + 
\mu_{\PV\PV,r(i)} N^{4\ell}_{\ZZ,i}(\theta) \Big),
\\
\mathcal{L} _{\Pe\PGm}=&\mathcal{P} \Big( N^{\Pe\PGm}_{\text{obs}} \mid \mu_{\text{NRB}}N^{\Pe\PGm}_{\text{NRB}}(\theta) + 
N^{\Pe\PGm}_{\text{other}}(\theta) \Big),
\\
\begin{split}
\mathcal{L} _{\text{DY}} =&  \mathcal{P} \Big( N^{\text{DY}}_{\text{obs}} \mid 
\mu_{\text{DY}}N^{\text{DY}}_{\text{DY}}(\theta) +\mu_{\text{NRB}}N^{\text{DY}}_{\text{NRB}}(\theta) 
\\&+N^{\text{DY}}_{\text{other}}(\theta) + N^{\text{DY}}_{\ZZ}(\theta) + N^{\text{DY}}_{\WZ}(\theta) + 
\mu N^{\text{DY}}_{\text{Sig}}(\theta) \Big).
\end{split}
\end{align}
The purpose of the fit is to determine the confidence interval for the signal strengths  $\mu$. Here $\mathcal{P}(N\mid \lambda)$ is the Poisson probability to observe $N$ events for an expected value of $\lambda$, 
and $f_{\text{NP}}(\theta \mid \hat{\theta})$ describes the nuisance parameters with log-normal probability density functions used for normalization uncertainties and Gaussian functions used for shape-based uncertainties.
The index $i$ indicates the bin of the \ptmiss or \mT distribution, 
 $r(i)$ corresponds to the region (low, medium, high) of bin $i$, and the index $j$ indicates either the 0-jet or 1-jet selection.
The diboson process normalization in the region $r(i)$ is $\mu_{\PV\PV,r(i)}$, while $\mu_{\text{DY}}$ is the DY background normalization and  
$\mu_{\text{NRB}}$ is the normalization for the nonresonant background.  
The yield prediction from simulation for process $x$ in region $y$ is noted as 
$N^{y}_{x}$. The smaller backgrounds in each region are merged together and are indicated collectively as "other". 
The method above for constructing likelihood functions follows 
that of Ref.~\cite{LHC-HCG}, where a more detailed mathematical description may be found.

\section{Systematic uncertainties} \label{sec:systematics}

In the following, we describe all of the uncertainties that are taken into account in the maximum likelihood fit.
We consider the systematic effects on both the overall normalization  
and on the shape of the distribution of  \ptmiss or \mT for all applicable uncertainties. 
We evaluate the impacts by performing the full 
analysis with the value of the relevant parameters shifted up and down by one standard deviation. The final varied 
distributions of \ptmiss or \mT  are used for signal extraction and as input to the fit. For each source of uncertainty, 
variations in the distributions are thus treated as fully correlated, while 
independent sources of uncertainty are treated as uncorrelated.  Except where noted otherwise, the systematic
uncertainties for the three different years of data taking are treated as correlated.

The assigned uncertainties in the integrated luminosity are 2.5, 2.3, and 2.5\% for the 
2016, 2017, and 2018 data 
samples~\cite{CMS-PAS-LUM-17-001,CMS-PAS-LUM-17-004,CMS-PAS-LUM-18-002}, respectively, 
and are treated as uncorrelated across the different years.

We apply scale factors to all simulated samples to correct for discrepancies in the lepton reconstruction and 
identification efficiencies between data and simulation.  These factors are measured
using DY events in the $\PZ$ boson peak region~\cite{Sirunyan:2018fpa,Khachatryan:2015hwa,Sirunyan:2019bzr} that 
are recorded with unbiased triggers.  The factors depend on the lepton \pt and $\eta$ 
and are within a few percent of unity for electrons and muons. The uncertainty in the determination of the trigger 
efficiency leads to an uncertainty smaller than $1\%$ in the expected signal yield. 

For the kinematic regions used in this analysis, the lepton momentum scale uncertainty for 
both electrons and muons is well represented by a constant value of $0.5\%$.
The uncertainty in the calibration of the jet energy scale (JES) and resolution directly affects  the \ptmiss computation 
and all the selection requirements related to jets. The estimate of the 
JES uncertainty is performed by varying the JES. 
 The variation corresponds to a re-scaling of the jet 
four-momentum as 
$p \to p (1 \pm \delta\pt^{\text{JES}}/\pt)$, where $\delta\pt^{\text{JES}}$ is 
the absolute uncertainty in the JES, which is parameterized as function of the \pt and $\eta$ 
of the jet. In order to account for the systematic uncertainty from the jet resolution smearing procedure, 
the resolution scale factors are varied within their uncertainties.
Since the uncertainties in the JES are derived independently for the three data sets, 
they are treated as uncorrelated across the three data sets.

The signal processes
are expected to produce very few events containing \PQb jets, and we reject events with 
any jets that satisfy the \PQb tagging algorithm working point used.
In order to account for the \PQb tagging efficiencies observed in data, an event-by-event reweighting using \PQb tagging scale 
factors and efficiencies is applied to simulated events. The uncertainty is 
obtained by varying the event-by-event weight by $\pm$1 standard deviation.
Since the uncertainties in the \cPqb tagging are derived independently for the three data sets, 
they are treated as uncorrelated across the three data sets. The variation of
the final yields induced by this procedure is less than $1\%$.

Simulated samples are reweighted to reproduce the pileup conditions observed in data. 
We evaluate the uncertainty related to pileup by recalculating these weights for variations 
in the total inelastic cross section by 5\% around the nominal value~\cite{Sirunyan:2018nqx}. 
The resulting shift in weights is propagated through the analysis 
and the corresponding \ptmiss and \mT spectra are used as input to the maximum likelihood fit. The variation of 
the final yields induced by this procedure is less than 1\%.

Shape-based uncertainties for the \ZZ and  \WZ backgrounds, referred to jointly as $\PV\PV$, and signal processes are derived from variations of the renormalization and factorization scales,
 the strong coupling constant \alpS, and PDFs~\cite{Rojo:2015acz,Butterworth:2015oua,Accardi_2016}.  
The scales are varied  up and down by a factor of two.
Variations of the PDF set and \alpS are used to estimate the corresponding uncertainties in 
the yields of the signal and background processes following Ref.~\cite{Ball:2017nwa}.
The missing higher-order EW terms in the event generation for the 
$\PV\PV$ processes yield another source of theoretical uncertainty~\cite{Bierweiler:2013dja,Gieseke:2014gka}. 
The following additional higher-order corrections are applied: a constant (approximately $10\%$) correction for the 
\WZ cross section from NLO to NNLO in QCD calculations~\cite{Grazzini:2016swo}; a constant (approximately $3\%$) correction 
for the \WZ cross section from LO to NLO in EW calculations, according to Ref.~\cite{Baglio:2013toa}; 
a $\Delta\phi(\PZ, \PZ)$-dependent correction to the \ZZ production cross section from NLO to next-to-next-to-leading order (NNLO) 
in QCD calculations~\cite{Grazzini:2015hta}; a $p_\mathrm{T}$-dependent correction to the \ZZ cross section from LO to NLO in EW calculations, 
following Refs.~\cite{Baglio:2013toa, Bierweiler:2013dja, Gieseke:2014gka}, which is the dominant correction in the signal region. 
We use the product of the above NLO EW corrections and the inclusive NLO QCD corrections~\cite{Campbell:2011bn} as an estimate of 
the missing NLO EW$\times$NLO QCD contribution, which is not used as a correction, but rather assigned as an uncertainty.
The resulting variations in the \ptmiss and \mT distribution are used as a shape uncertainty in the likelihood fit.

The shapes of the \ptmiss and \mT distributions 
are needed for each of the background processes.
For the DY and nonresonant processes, we take the shape directly from simulation. 
The distributions for the \ZZ and \WZ processes are obtained by taking the shapes from the 
simulation and normalizing them to the yield seen in the data in the CR.
The gluon-induced and the quark-induced \ZZ processes have different acceptances and their uncertainties are treated separately, while the normalization factors are taken to be correlated. 
In all cases, the limited number of simulated events in 
any given bin gives rise to a systematic uncertainty. This uncertainty is treated as fully uncorrelated across the bins and processes.

A summary of the impact on the signal strength of the systematic uncertainties is shown in Table~\ref{tab:systematics}.  The 
 $\PZ\Ph(\text{invisible})$ model is used as an example to illustrate the size of the uncertainties, 
 both for the presence ($\mathcal{B}(\Ph \to \text{invisible})=1$) and absence 
 ($\mathcal{B}(\Ph \to \text{invisible})=0$) of a signal. 
These two paradigms are used to generate Asimov data sets that are then fit to give the uncertainty estimates shown in Table~\ref{tab:systematics}. 
The systematic uncertainties are dominated by the 
theoretical uncertainty in the \ZZ and \WZ background contributions.

\begin{table*}[htbp]
\centering
\topcaption{
Summary of the uncertainties in the branching fraction arising from the systematic uncertainties considered in the $\PZ\Ph(\text{invisible})$ model 
assuming $\mathcal{B}(\Ph \to \text{invisible})=1$ (signal) 
and $\mathcal{B}(\Ph \to \text{invisible})=0$ (no signal). 
Here, lepton measurement refers to the combined trigger, lepton reconstruction and identification efficiencies, 
and the lepton momentum and electron energy scale systematic uncertainty. Theory uncertainties include variations of the 
renormalization and factorization scales, $\alpha_{s}$, and PDFs as well as the higher-order EWK corrections. \label{tab:systematics}} 
{
\begin{tabular}{lcc}
\hline
Source of uncertainty                        & Impact assuming signal  & Impact assuming no signal   \\
\hline
Integrated luminosity                                   &  \multirow{1}{*}{0.013}	   &  \multirow{1}{*}{0.002}	 \\
Lepton measurement                          &  \multirow{1}{*}{0.032}       &  \multirow{1}{*}{0.050}       \\
Jet energy scale and resolution		     &  \multirow{1}{*}{0.042}       &  \multirow{1}{*}{0.024}       \\
Pileup  				     &  \multirow{1}{*}{0.012}       &  \multirow{1}{*}{0.009}       \\
\PQb tagging efficiency		             &  \multirow{1}{*}{0.004}       &  \multirow{1}{*}{0.002}       \\
Theory            			     &  \multirow{1}{*}{0.088}       &  \multirow{1}{*}{0.085}       \\
Simulation sample size                          &  \multirow{1}{*}{0.024}       &  \multirow{1}{*}{0.023}	 \\ [\cmsTabSkip]
Total systematic uncertainty                 &  0.11    		   &  0.11			 \\
Statistical uncertainty                      &   0.089     		   &   0.073			 \\ [\cmsTabSkip]
Total uncertainty                            &  0.14     		   &  0.13		 \\
\hline
\end{tabular}
}
\end{table*}

\section{Results}\label{sec:results}

\begin{figure}[!hbtp]
  \centering
    \includegraphics[width=0.48\textwidth]{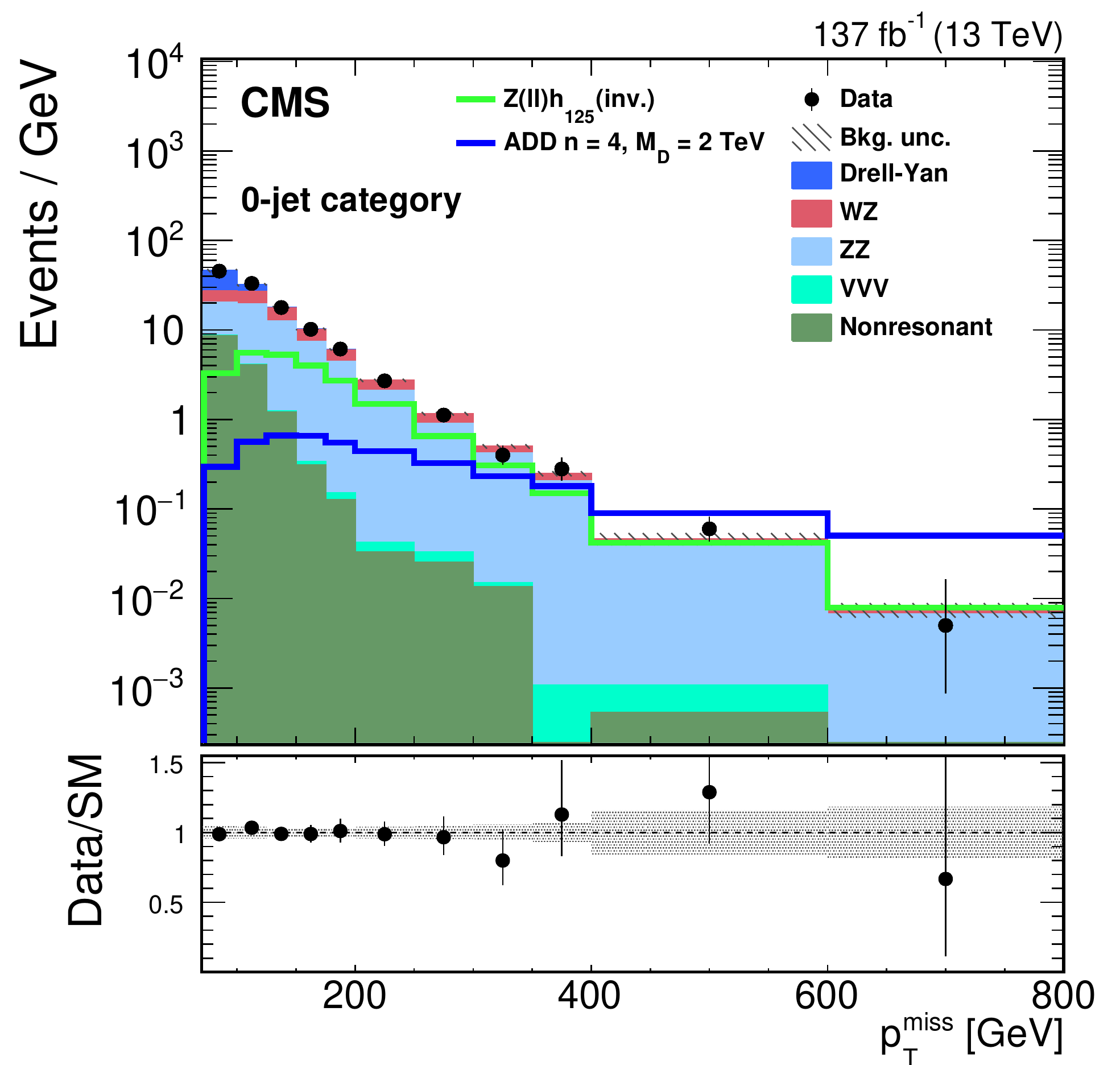}
    \includegraphics[width=0.48\textwidth]{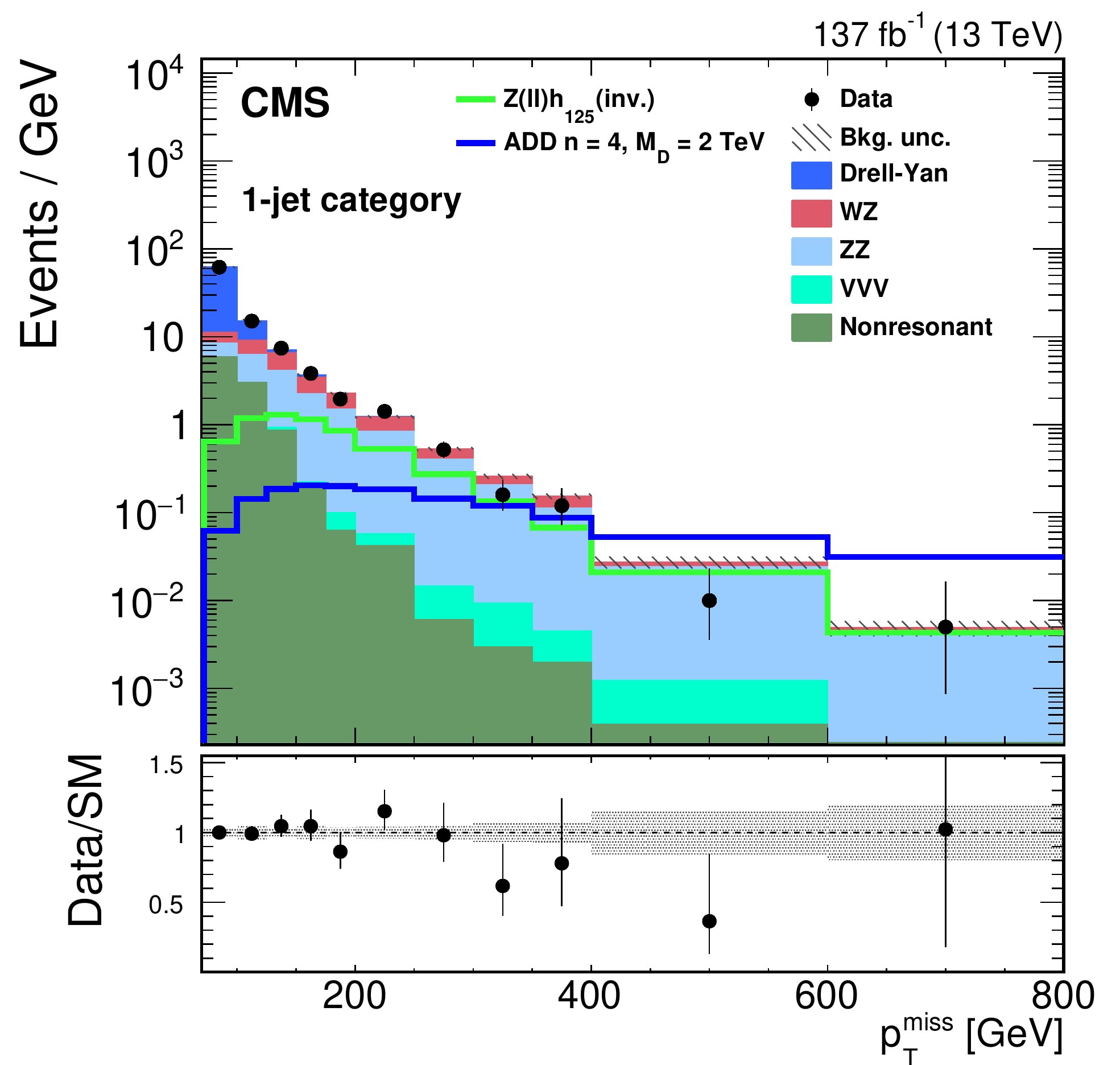}
  \caption{The \ptmiss distributions for events in the signal region in the 0-jet (\cmsLeft) and 1-jet (\cmsRight) categories. 
  The rightmost bin also includes events with $\ptmiss>800\GeV$. The uncertainty band includes both statistical and 
  systematic components. The $\PZ\Ph(\text{invisible})$ signal normalization assumes SM production rates 
  and the branching fraction $\mathcal{B}(\Ph \to \text{invisible})=1$. For the ADD model, the signal normalization 
  assumes the expected values for $n=4$ and $M_{\mathrm{D}}=2\TeV$.\label{fig:finallevel}
  }
\end{figure}

\begin{figure}[!hbtp]
  \centering
    \includegraphics[width=0.48\textwidth]{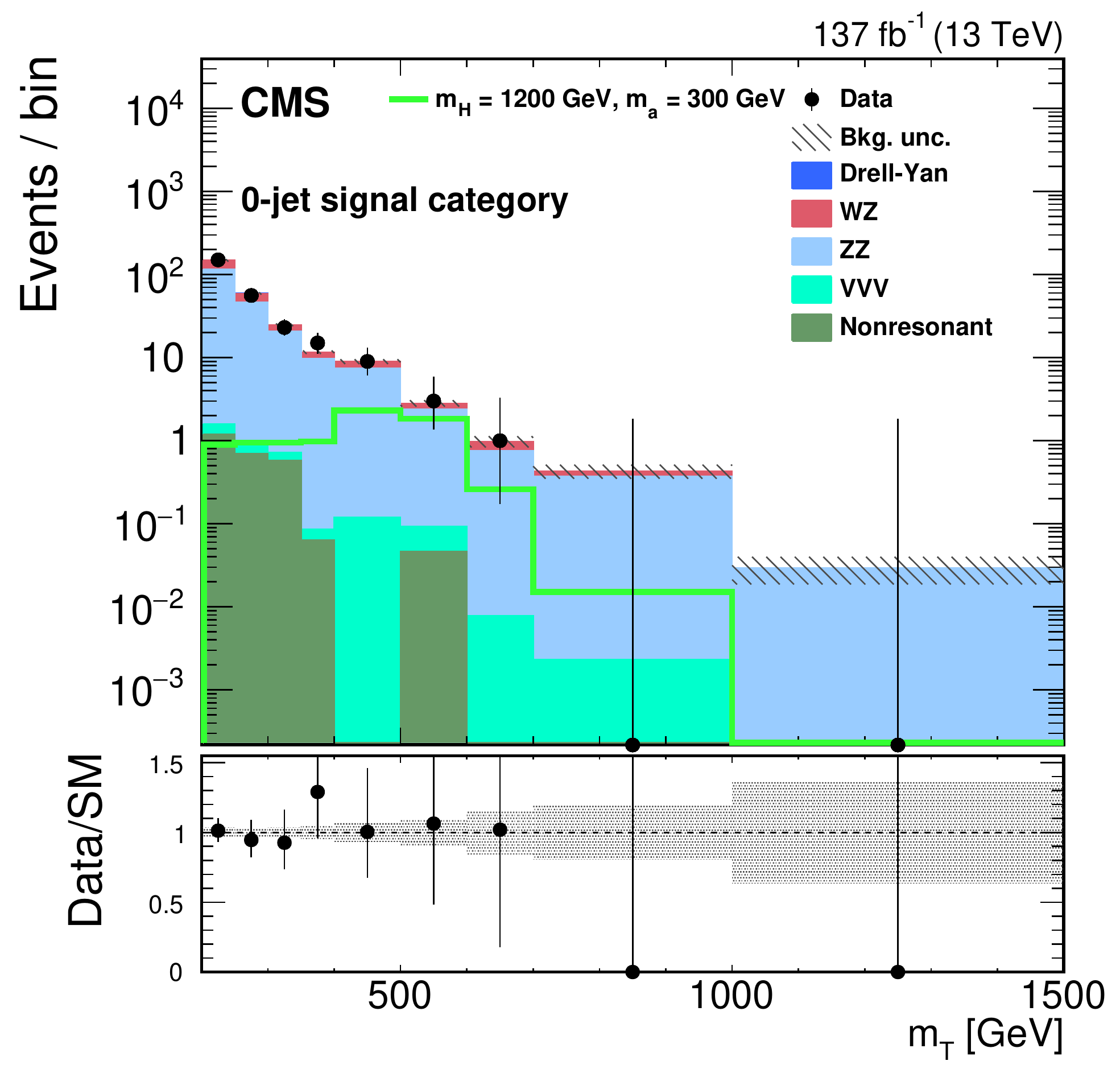}
    \includegraphics[width=0.48\textwidth]{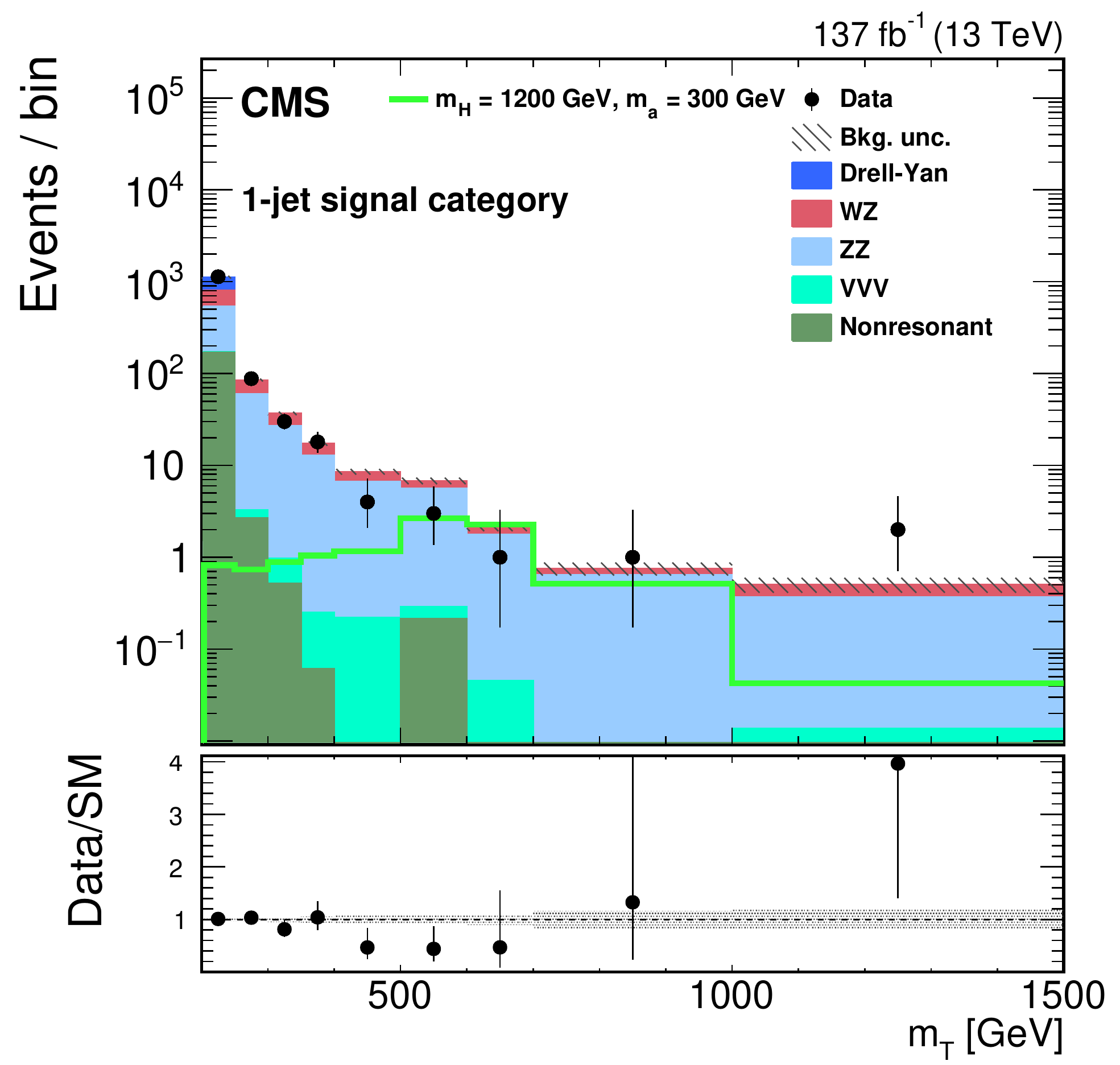}
  \caption{The \mT distributions for events in the signal region in the 0-jet (\cmsLeft) and 1-jet (\cmsRight) categories. 
  The rightmost bin includes all events with $\mT>1000\GeV$. The uncertainty band includes both statistical and 
  systematic components. The signal normalization assumes the expected values for $m_\PH=1200\GeV, m_{\aboson}=300\GeV$ within 
  the 2HDM+\aboson framework where $m_{\PH}=m_{\Hpm}=m_{\PSA}$, $\tan\beta=1$ and $\sin\theta=0.35$
\label{fig:finallevel_MT}
  }
\end{figure}

The number of observed and expected events in the SR after the final selection is given in 
Table~\ref{tab:zhinvsel}, where the values of the expected yields and their uncertainties are obtained from
the maximum likelihood fit. 
The observed numbers of events are compatible with the background predictions. 
The expected yields and the 
product of acceptance and efficiency for several signal models used in the analysis 
are shown in Table~\ref{tab:signsel}. 
The post-fit \ptmiss distributions for events in the signal region in the 0-jet and 1-jet categories 
 are shown in Fig.~\ref{fig:finallevel}. 
The final $\mT$ distributions used for the 2HDM+\aboson model are shown in Fig.~\ref{fig:finallevel_MT}. 

\setlength{\tabcolsep}{12pt}
\begin{table*}[hbtp]
  \topcaption{Observed number of events and post-fit background estimates
   in the two jet multiplicity categories of the SR. The reported uncertainty represents the sum in quadrature of the statistical and systematic components.
  \label{tab:zhinvsel}}
  \centering
{
  \begin{tabular}{lcc}
\hline 
Process & 0-jet category & 1-jet category \\
\hline
Drell-Yan               &   $502\pm94$ &  $1179\pm64$ \\ 
\WZ  &  $1479\pm53$ &   $389\pm16$ \\ 
\ZZ                &   $670\pm27$ &   $282\pm13$ \\
Nonresonant background      &   $384\pm31$ &   $263\pm22$ \\
Other background            &     $6.3\pm0.7$ &  $6.8\pm0.8$ \\ [\cmsTabSkip]
Total background            &  $3040\pm110$ &  $2120\pm76$ \\ [\cmsTabSkip]
Data                  &  3053 & 2142 \\
\hline
  \end{tabular}
}
\end{table*}

\setlength{\tabcolsep}{12pt}
\begin{table*}[hbtp]
  \topcaption{Expected yields and the product of acceptance and efficiency for several models probed in the analysis.
  The quoted values correspond to the $\PZ\to\ell\ell$ decays. The reported uncertainty represents the sum in quadrature of the statistical and systematic components.
  \label{tab:signsel}}
  \centering
{
  \begin{tabular}{lcc}
\hline
Model & Yields & Product of acceptance \\ & & and efficiency (\%) \\
\hline
$\PZ\Ph(125)$                           & $864\pm64$ & $10.6\pm0.8$ \\
ADD $M_{\mathrm{D}}=3\TeV, n=4$              &  $35.1\pm2.4$ & $18.6\pm1.3$ \\
Unparticle $S_\textsf{U}=0, d_\textsf{U}=1.50$      & $221\pm16$ &  $8.2\pm0.6$ \\
2HDM+\aboson $m_{\PH}=1000\GeV, m_{\aboson}=400\GeV$ &  $14.1\pm4.0$ & $12.7\pm2.7$ \\
DM Vector $m_{\text{med}}=1000\GeV, m_{\PGc}=1\GeV$           &  $64.8 \pm6.1$ & $17.6\pm1.7$ \\
\hline
  \end{tabular}
}
\end{table*}

For each of the models considered, simulated signal samples are generated for relevant sets of model parameters.  
The observed \ptmiss and $\mT$ spectra are used to set limits on theories of new physics using the modified frequentist construction \CLs~\cite{Read1,LHC-HCG,junkcls} used in the asymptotic approximation~\cite{pvalue}.

\subsection{Simplified dark matter model interpretation}

\begin{figure}[!hbtp]
  \centering
   \includegraphics[width=0.48\textwidth]{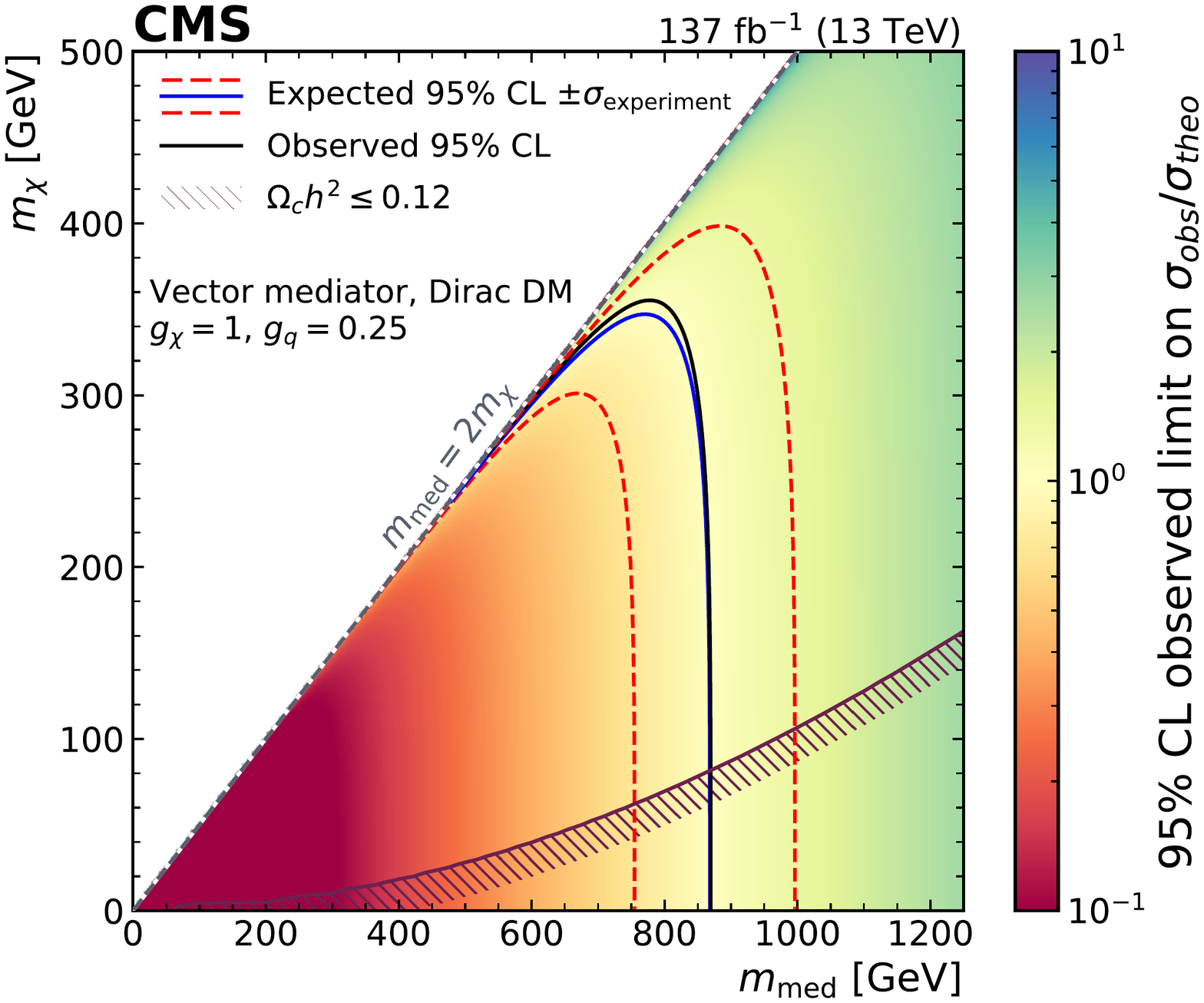}
   \includegraphics[width=0.48\textwidth]{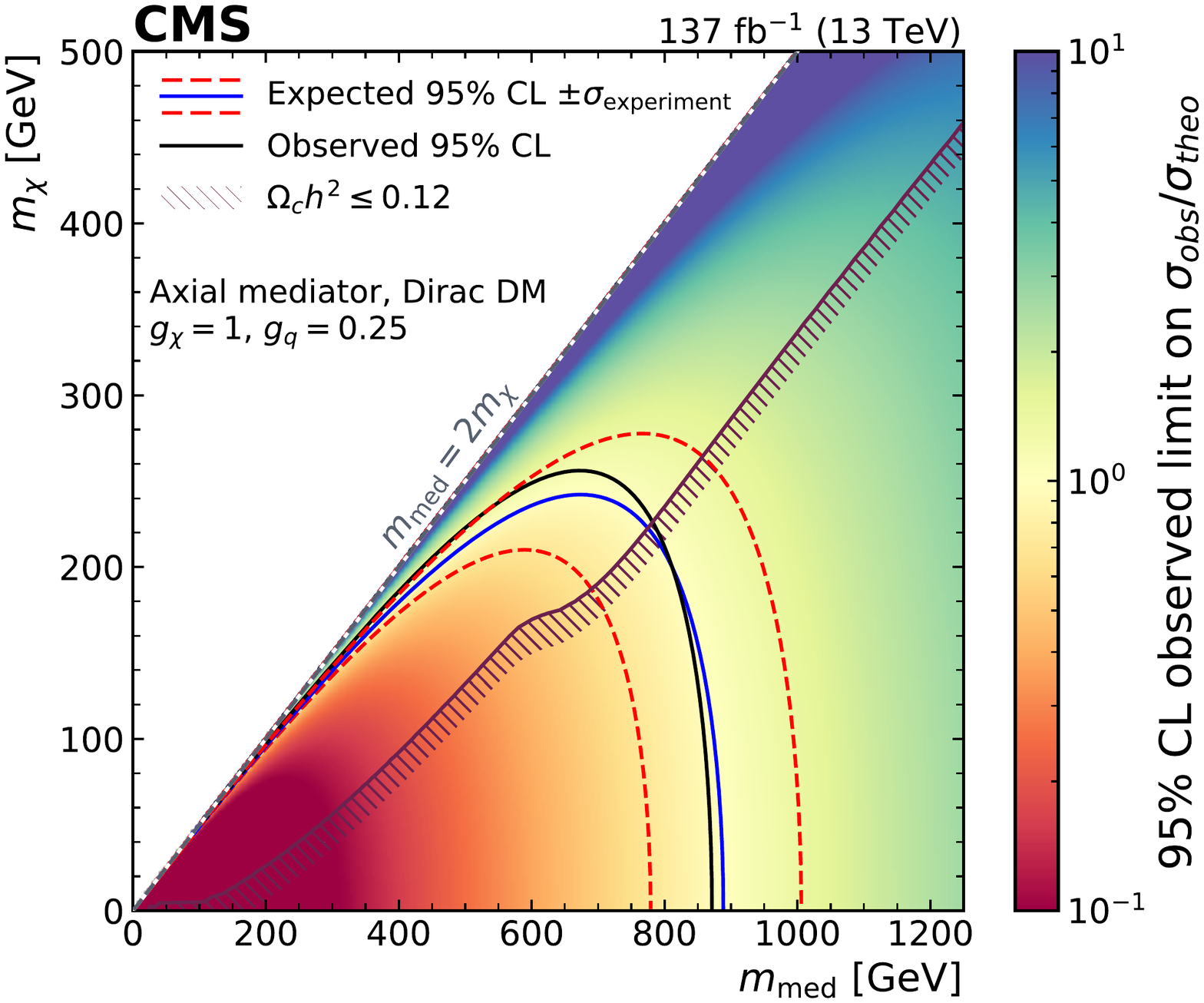}
  \caption{
The 95\% \CL exclusion limits for the vector (\cmsLeft) and the axial-vector (\cmsRight) simplified models. 
The limits are shown as a function of the mediator and DM particle masses. The coupling to quarks 
is fixed to $g_{\Pq}=0.25$ 
and the coupling to DM is set to $g_\chi=1$.
}
  \label{fig:VectorDMLimits}
\end{figure}

\begin{figure}[!hbtp]
  \centering
    \includegraphics[width=0.48\textwidth]{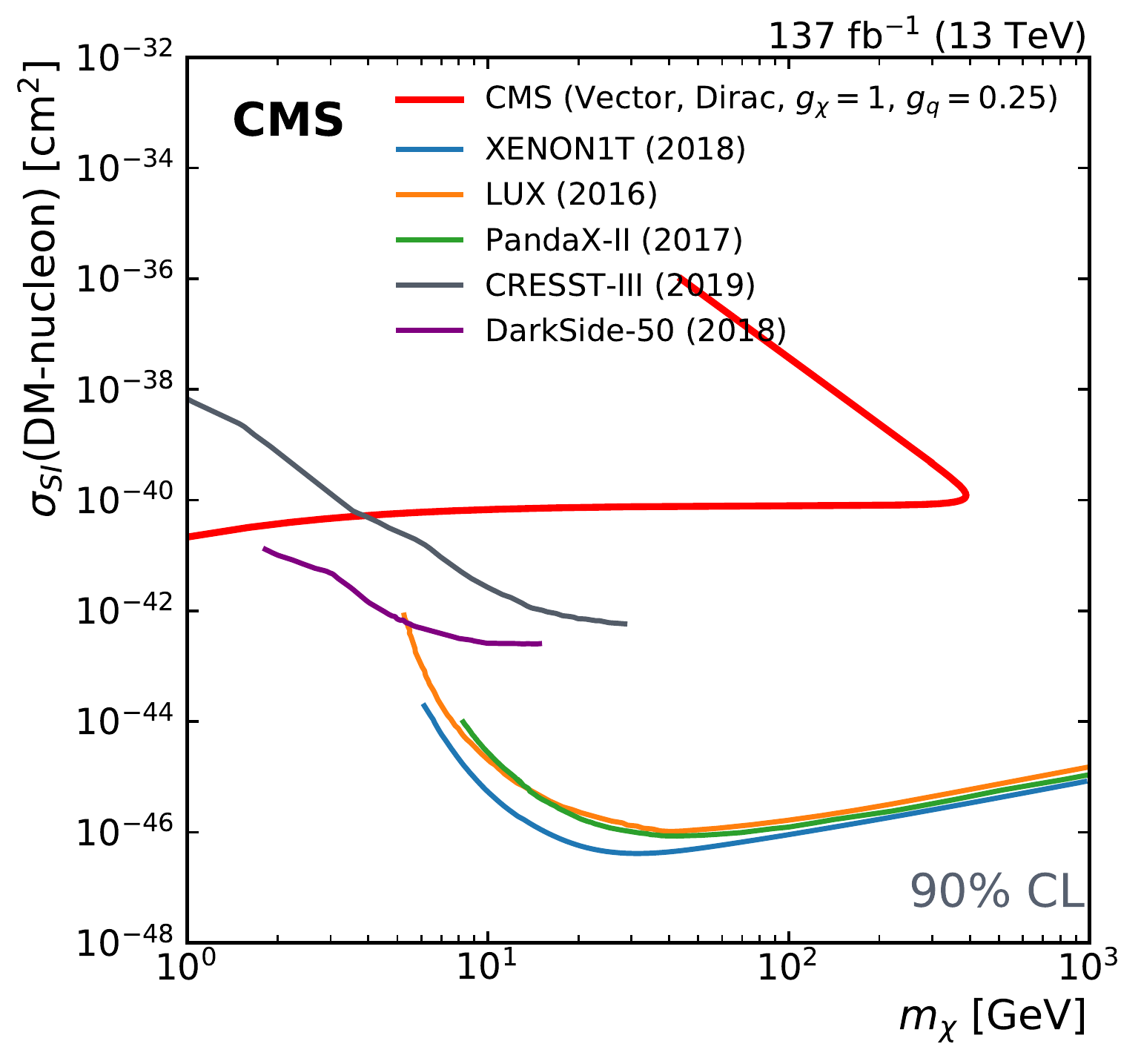}
    \includegraphics[width=0.48\textwidth]{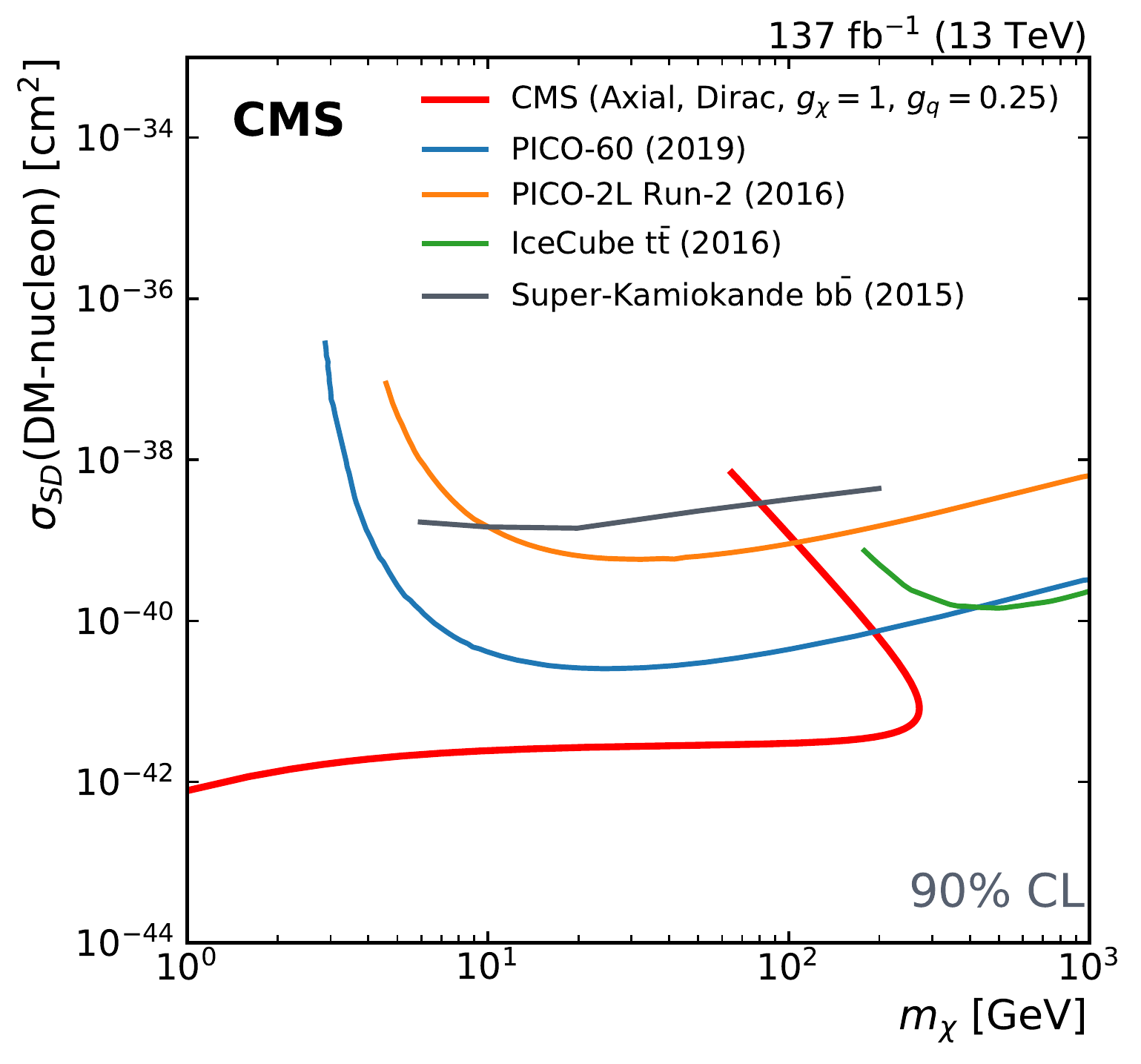}

  \caption{The 90\% \CL DM-nucleon upper limits on the cross section for simplified DM in the spin-independent (\cmsLeft) and spin-dependent (\cmsRight) cases. 
  The coupling to quarks is set to $g_{\Pq}=0.25$ and the coupling to DM is set to $g_\chi=1$. 
Limits from the XENON1T~\cite{Aprile_2018}, LUX~\cite{Akerib_2017}, PandaX-ll~\cite{Cui_2017}, CRESST-III~\cite{Abdelhameed:2019hmk}, and DarkSide-50~\cite{Agnes_2018} experiments are shown for the spin-independent case with vector couplings.
Limits from the PICO-60~\cite{Amole_2019}, PICO-2L~\cite{Amole:2016pye}, IceCube~\cite{Aartsen:2016exj}, and Super-Kamiokande~\cite{Choi:2015ara} experiments are shown for the spin-dependent case with axial-vector couplings.  
\label{fig:limits_DD}
  }
\end{figure}

In the framework of the simplified models of DM, the signal production is sensitive to the mass, spin, and 
parity of the mediator as well as the coupling strengths of the mediator to quarks and to DM. 
The \ptmiss distribution is used as an input to the fit. Limits for the vector and axial-vector 
mediators are shown as a function of the mediator mass $m_{\text{med}}$ and DM particle mass $m_\PGc$ as shown in 
Figure~\ref{fig:VectorDMLimits}. Cosmological constraints on the DM abundance~\cite{Albert:2017onk} 
are added to Fig.~\ref{fig:VectorDMLimits} where the shaded area 
represents the region where additional physics would be needed to describe the DM abundance. For vector mediators, we observe a limit around $m_{\text{med}}>870\GeV$ for most values 
of $m_\PGc$ less than $m_{\text{med}}/2$.  For axial-vector mediators 
the highest limit reached in the allowed region is about $m_{\text{med}}>800\GeV$.  In both cases, the previous limits from this 
channel are extended by about 150\GeV, but the limits are still less restrictive than those from 
published mono-jet results~\cite{Sirunyan:2017jix} because weakly coupled \PZ bosons are radiated from the initial state 
quarks much less frequently than gluons. 
Figure~\ref{fig:limits_DD} shows the 90\% \CL limits on the DM-nucleon cross sections calculated following the suggestions in Ref.~\cite{Boveia:2016mrp}. 
Limits are shown as a function of the DM particle mass for both the spin-independent and spin-dependent cases and compared to selected 
results from direct-detection experiments. 

\begin{figure}[!hbtp]
  \centering
    \includegraphics[width=0.48\textwidth]{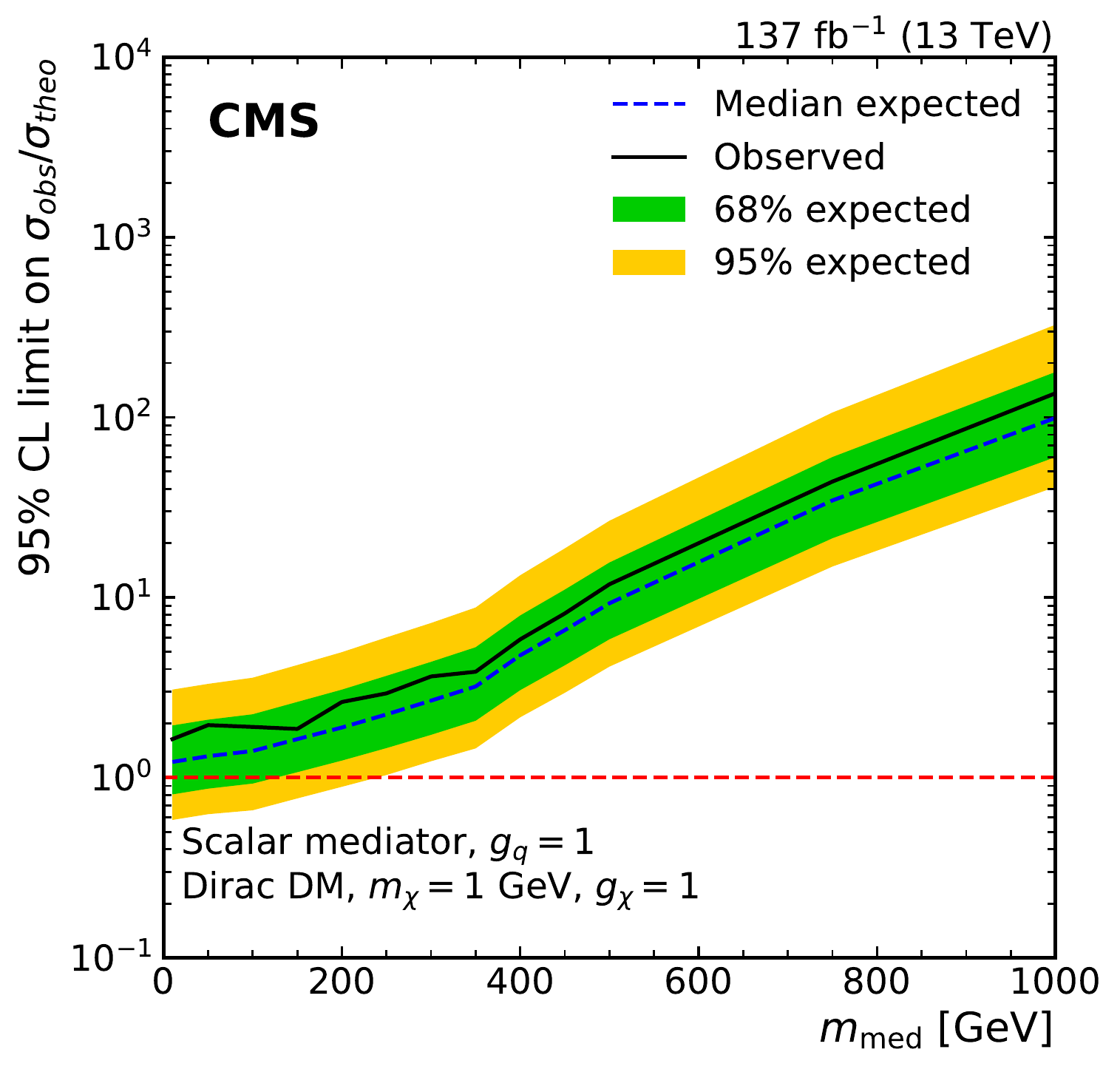}
    \includegraphics[width=0.48\textwidth]{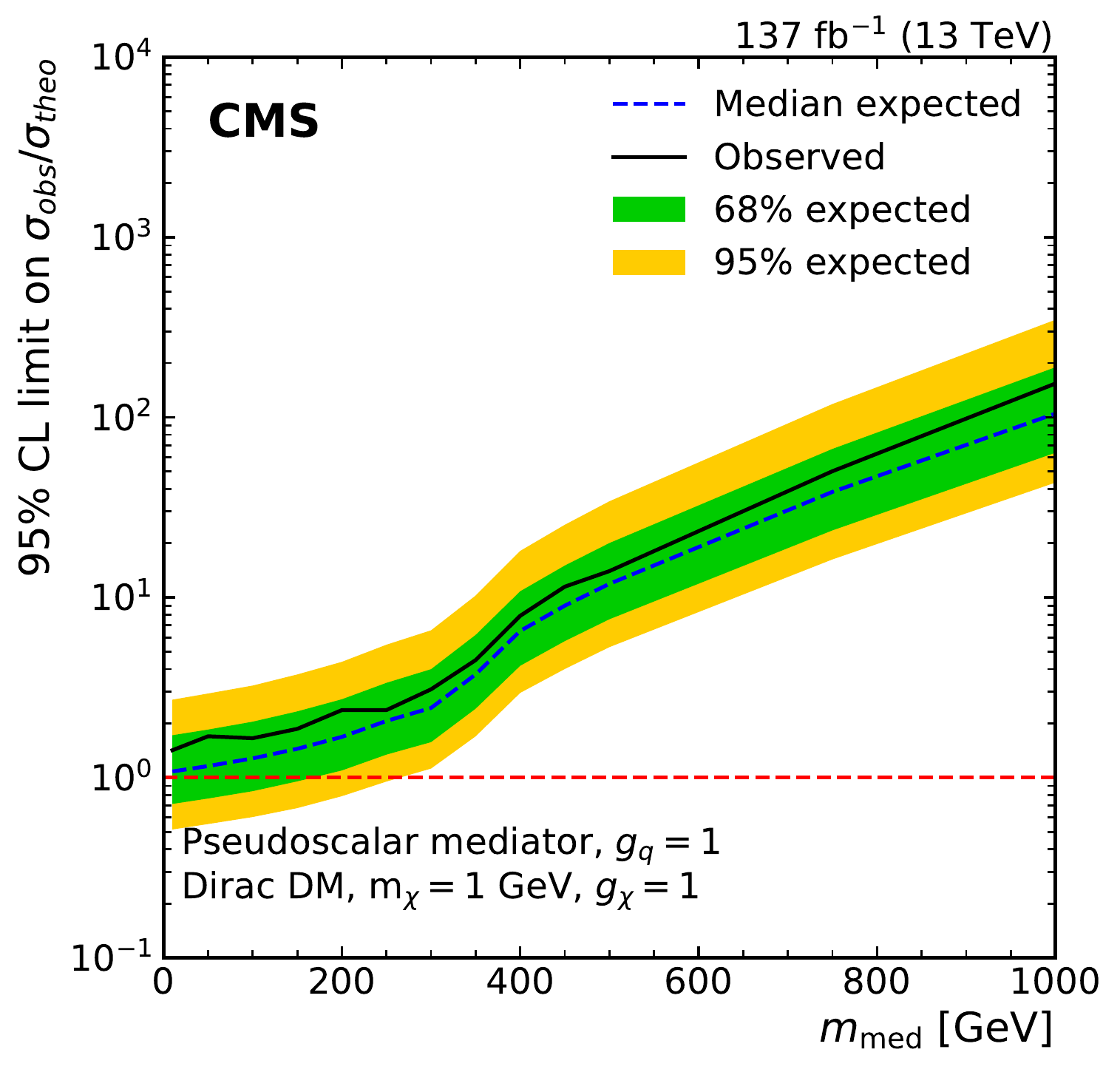}

  \caption{The 95\% \CL upper limits on the cross section for simplified DM models with scalar (\cmsLeft) and pseudoscalar (\cmsRight) mediators. 
  The coupling to quarks is set to $g_{\Pq}=1$, the 
coupling to DM is set to $g_\chi=1$ and the DM mass is  $m_\chi=1\GeV$.\label{fig:limits_DM_scalar}  
  }
\end{figure}

In addition to vector and axial-vector mediators, scalar and pseudoscalar mediators are also tested. For these models, we fix
both couplings to quarks and to DM particles: $g_{\Pq}=1$ and  $g_\chi=1$ as suggested in Ref.~\cite{Boveia:2016mrp}.
Since the choice of DM particle mass is shown to have negligible effects on the kinematic distributions of the detected particles, 
we set it to the constant value of $m_\PGc=1\GeV$. 
 Figure~\ref{fig:limits_DM_scalar} gives the 95\% \CL exclusion limits on the production cross section over the predicted cross 
 section as a function the mediator mass $m_{\text{med}}$.  The expected limits are about 25\% better than the previous results in this channel~\cite{Sirunyan:2017qfc}, 
but are not yet sensitive enough to exclude any value of $m_{\text{med}}$.  The best limits obtained on the cross section are about 
1.5 times larger than the predicted values for low values of $m_{\text{med}}$.

\subsection{Two-Higgs-doublet model interpretation}

\begin{figure}[!hbtp]
  \centering
   \includegraphics[width=0.48\textwidth]{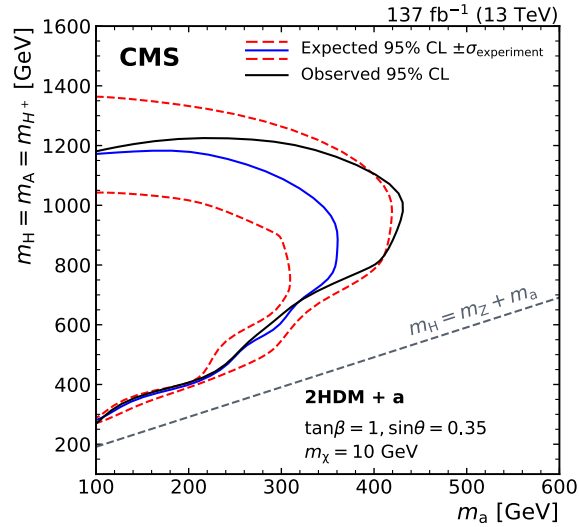}
  \caption{
The 95\% \CL upper limits on the 2HDM+\aboson model with the mixing angles set to $\tan\beta=1$ and $\sin\theta=0.35$ and with a 
DM particle mass of $m_{\PGc}=10\GeV$. The limits are shown as a function of the heavy Higgs boson and the pseudoscalar masses.}
  \label{fig:2HDMLimits}
\end{figure}

For the 2HDM+\aboson model, the signal production is sensitive to the heavy Higgs boson and the pseudoscalar \aboson masses. 
As discussed in Section~\ref{sec:fitting}, the \mT distribution is used in the fit rather than \ptmiss. 
The limits on both the heavy Higgs boson and the additional pseudoscalar mediator \aboson are shown in Fig.~\ref{fig:2HDMLimits}. 
The mixing angles are set to $\tan\beta=1$ and $\sin\theta=0.35$ with a DM particle mass of $m_{\PGc}=10\GeV$.
The mediator mass with the most sensitivity is $m_{\PH}=1000\GeV$, 
where the observed (expected) limit on $m_{\aboson}$ is 440 (340)\GeV. 
For small values of $m_{\aboson}$, the limit on $m_\PH$ is about 1200\GeV. 
These can be compared with the observed (expected) limits from ATLAS of 
$m_{\aboson}>340$ (340)\GeV and $m_{\PH}>1050$ (1000)\GeV based on a $\sqrt{s}=13\TeV$ data set corresponding to an 
integrated luminosity of 36\fbinv~\cite{Aaboud:2019yqu}.   

\subsection{Invisible Higgs boson interpretation}
\begin{figure}[!hbtp]
  \centering
   \includegraphics[width=0.48\textwidth]{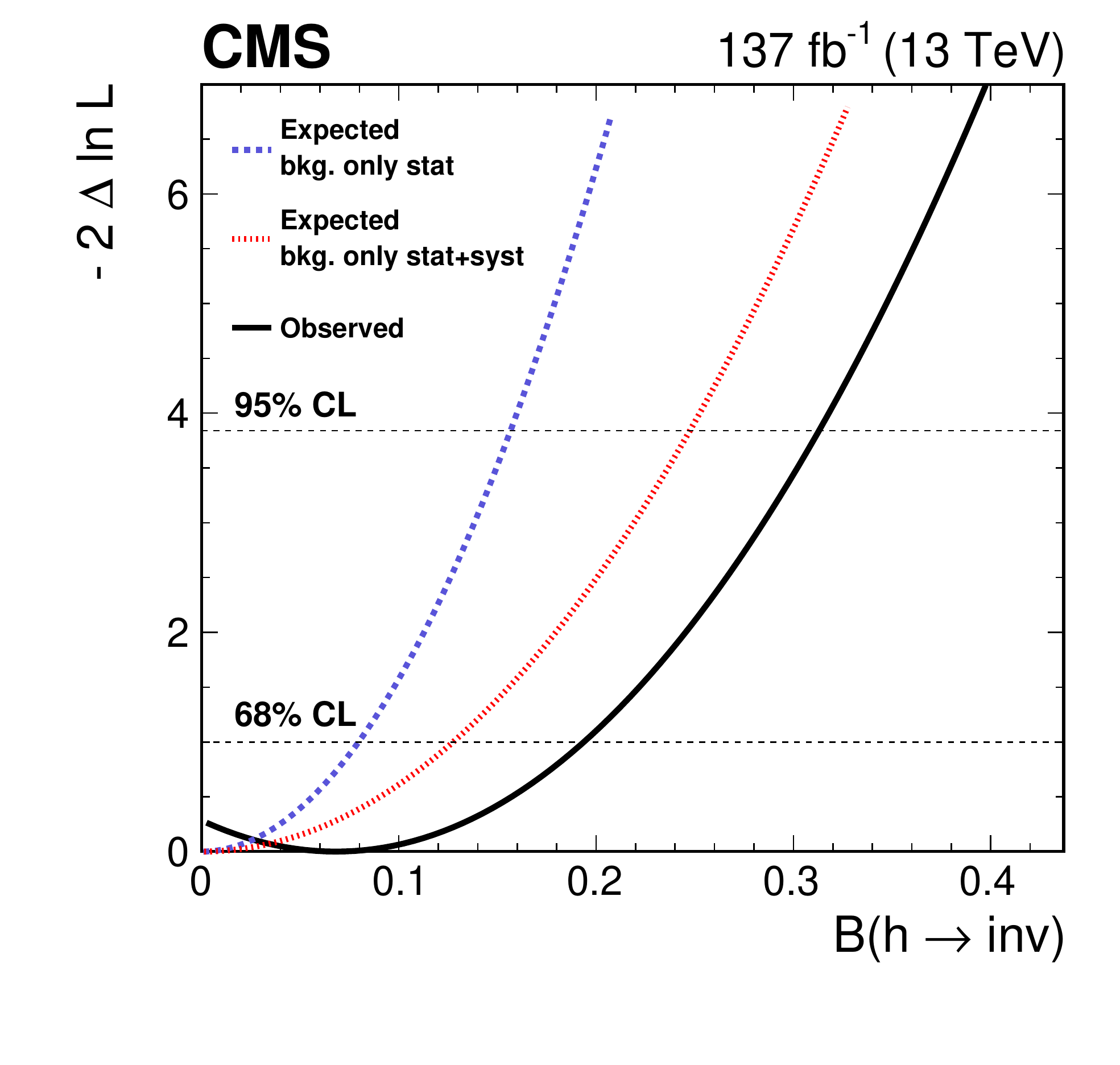}
  \caption{
The value of the negative log-likelihood, $-2\Delta$ln$\mathcal{L}$, as a function of the branching fraction of the Higgs boson decaying to invisible particles.}
  \label{fig:Zh_Likelihood}
\end{figure}

For the search for invisible decays of the Higgs boson, we use the \ptmiss distribution as input to the fit.  We obtain
upper limits on the product of the Higgs boson production cross section 
and branching fraction to invisible particles
$\sigma_{\PZ\Ph}\mathcal{B}(\Ph \to {\text{invisible}})$.
This can be interpreted as an upper limit on $\mathcal{B}(\Ph \to {\text{invisible}})$ 
by assuming the production rate~\cite{Heinemeyer:2013tqa,deFlorian:2016spz,Harlander:2018yio} for 
an SM Higgs boson at $m_{\Ph} = 125\GeV$. 
The observed (expected) 95\% \CL upper limit at $m_{\Ph} = 125\GeV$ on 
$\mathcal{B}(\Ph \to {\text{invisible}})$ is 29\% ($25^{+9}_{-7}$\%) as shown in Fig.~\ref{fig:Zh_Likelihood}.  The observed (expected) limit from the previous CMS result in this channel was $\mathcal{B}(\Ph \to {\text{invisible}})< 45 (44)$\%.  The combinations of all earlier results
yields an observed (expected) limit of 19 (15)\% from CMS~\cite{Sirunyan:2018owy} and 26\% ($17^{+5}_{-5}$\%) from ATLAS~\cite{Aaboud:2019rtt}.

\subsection{Unparticle interpretation}
\begin{figure}[!hbtp]
  \centering
   \includegraphics[width=0.48\textwidth]{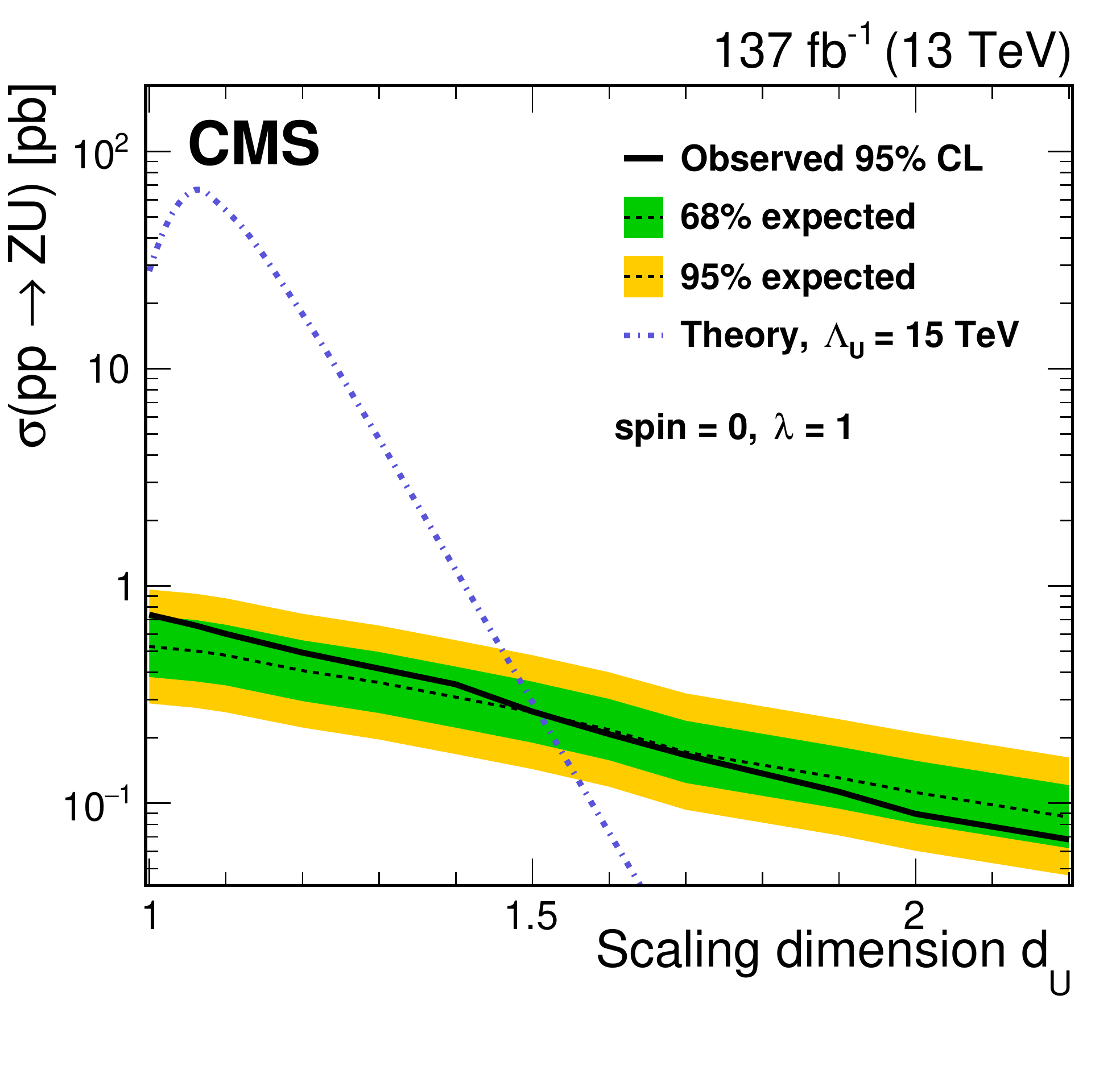}
  \caption{
The 95\% \CL upper limits on unparticle+\PZ production cross section, as a function of the scaling dimension $d_\textsf{U}$.
These limits apply to fixed values of the effective cutoff scale $\Lambda_\textsf{U}=15\TeV$ 
and coupling $\lambda=1$.} 
  \label{fig:unparticleLimits}
\end{figure}

In the unparticle scenario, the same analysis of the \ptmiss spectrum is performed.
At 95\% \CL, upper limits are set on the cross section with $\Lambda_\textsf{U}=15\TeV$. 
The limits are shown in Fig.~\ref{fig:unparticleLimits} as a function of the scaling dimension $d_\textsf{U}$.
The observed (expected) limits are 0.5 (0.7)\unit{pb}, 0.24 (0.26)\unit{pb}, and 0.09 (0.07)\unit{pb} for $d_\textsf{U} = 1$, $d_\textsf{U} = 1.5$, and $d_\textsf{U} = 2$ respectively, compared to
1.0 (1.0)\unit{pb}, 0.4 (0.4)\unit{pb}, and 0.15 (0.15)\unit{pb} for the earlier result~\cite{Sirunyan:2017qfc}. These limits depend on the choice 
of $\lambda$ and $\Lambda_\textsf{U}$, 
as the cross section scales with the Wilson coefficient $\lambda/\Lambda_\textsf{U}$~\cite{Georgi:2007ek}. 
We fix the coupling between the SM and the unparticle fields to $\lambda=1$.

 \subsection{The ADD Interpretation}
\begin{figure}[!hbtp]
  \centering
    \includegraphics[width=0.48\textwidth]{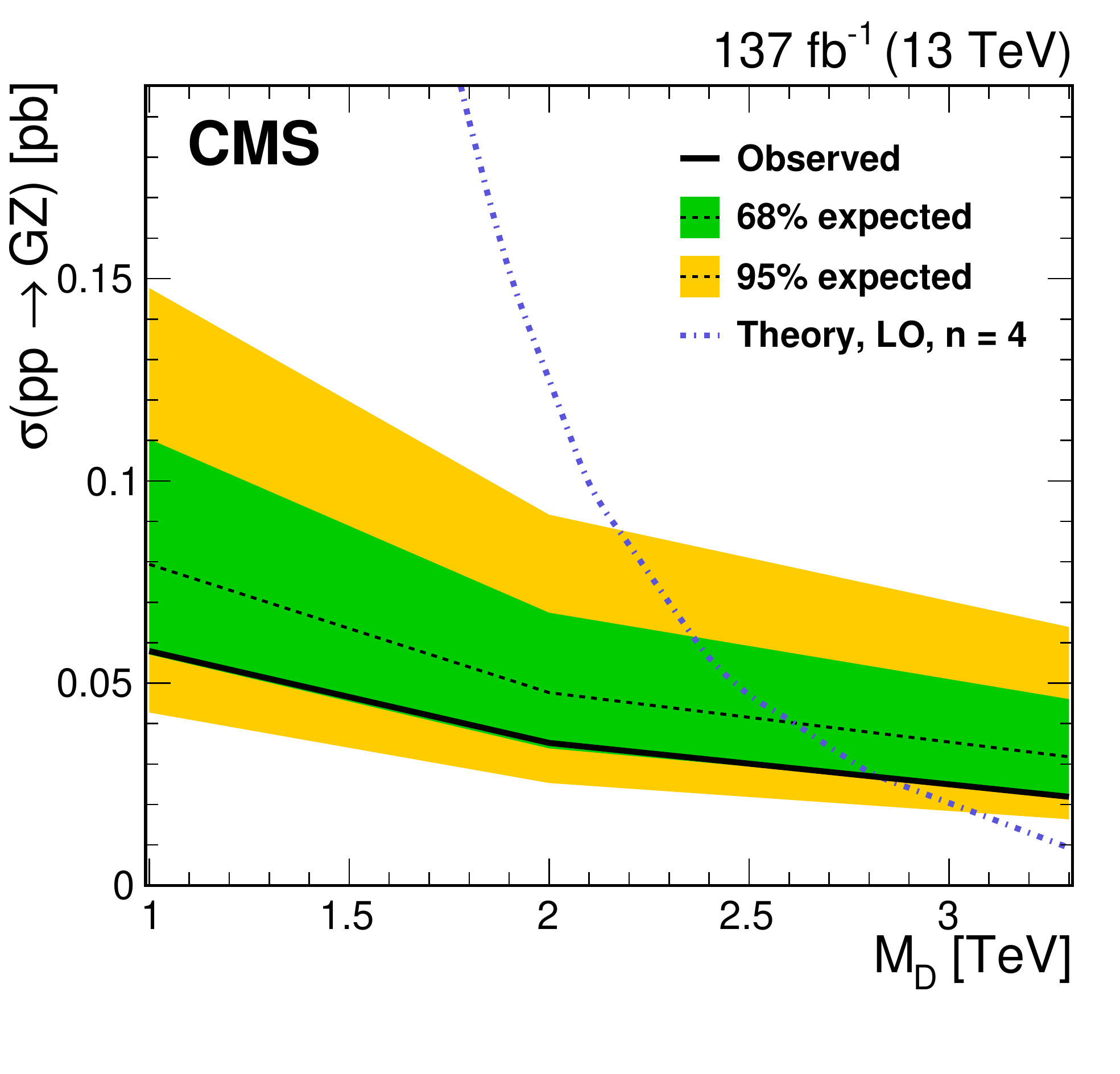}

  \caption{The 95\% \CL cross section limit in the ADD scenario as a function of $M_{\mathrm{D}}$ for $n=4$.\label{fig:limits_add_xs}
  }
\end{figure}

\begin{figure}[!hbtp]
  \centering
  \includegraphics[width=0.48\textwidth]{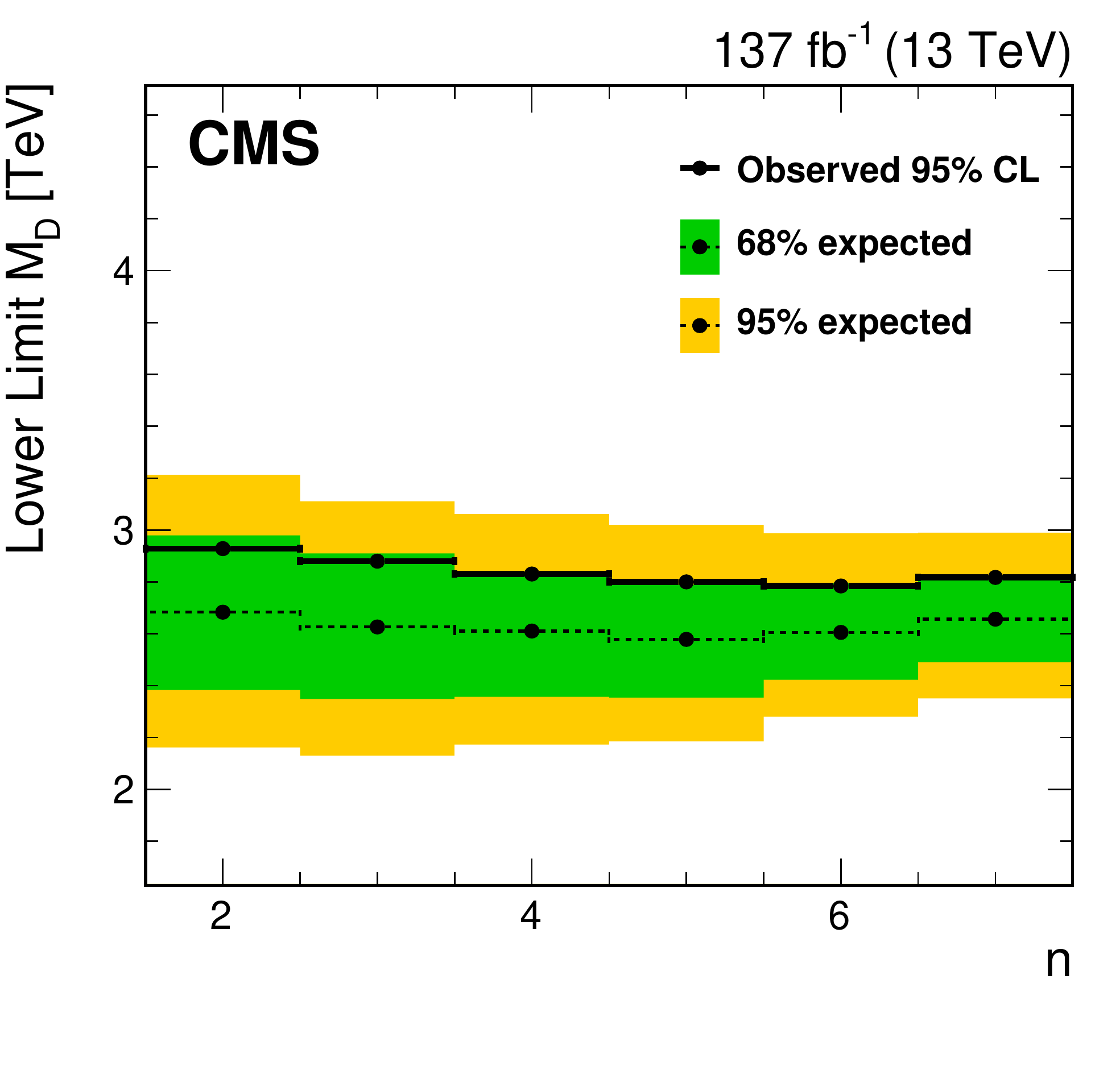}
  \caption{The 95\% \CL expected and observed exclusion limits on $M_{\mathrm{D}}$ as a function of the number of
  extra dimensions $n$.\label{fig:limits_add_md}
  }
\end{figure}

In the framework of the ADD model of extra dimensions, we use the fits to the  \ptmiss distribution to calculate limits on the number 
of extra dimensions $n$ and the fundamental Planck scale $M_{\mathrm{D}}$. 
The cross section limit calculated as a function of $M_{\mathrm{D}}$ for the case where $n=4$ is shown in Fig.~\ref{fig:limits_add_xs}.
The limits on $M_{\mathrm{D}}$ as a function of $n$ are obtained, as  
shown in Fig.~\ref{fig:limits_add_md}. The observed (expected) 95\% \CL exclusion upper limit on the 
mass $M_{\mathrm{D}}$ is 2.9--3.0 (2.7--2.8)\TeV 
compared to earlier results of 2.3--2.5 (2.3--2.5)\TeV~\cite{Sirunyan:2017qfc}.  

\subsection{Summary of limits}

Table~\ref{tab:201results} gives a summary of the limits expected and observed for a selection of relevant parameters in all of the models considered.

\begin{table*}[hbtp]
  \topcaption{Observed and expected 95\% \CL limits on parameters for the simplified DM models, invisible decays of the
Higgs boson, two-Higgs-doublet model, large extra dimensions in the ADD scenario, and unparticle model. For the scalar
and pseudoscalar mediators, the limits are dependent on the mediator mass, so the lowest values for the ratio of observed
to theoretical cross sections are presented. For the vector and axial-vector mediators, the limits are dependent on the DM
particle mass, so the limits are shown for $m_\PGc<300\GeV$ for the vector mediator and $m_\PGc=240\GeV$ for the axial-vector mediator. }
  \label{tab:201results}
  \centering
{
  \begin{tabular}{lccc}
\hline
Model & Parameter & Observed & Expected \\
\hline
DM - vector & $m_{\text{med}}$   & 870\GeV & 870\GeV \\
 $g_\chi=1$                &                      &         &        \\
 $g_{\Pq}=0.25$                &                      &         &        \\ [\cmsTabSkip]
DM - axial-vector& $m_{\text{med}}$   &  800\GeV & 800\GeV \\
 $g_\chi=1$                &                      &         &        \\
 $g_{\Pq}=0.25$                &                      &         &        \\ [\cmsTabSkip]
DM - scalar & $\sigma_\text{obs}/\sigma_\text{theo}$ &  1.8 & 1.5 \\
 $g_\chi=1$                &                      &         &        \\
 $g_{\Pq}=1$                &                      &         &        \\ 
 $m_\PGc=1\GeV$                &                      &         &        \\ [\cmsTabSkip]
DM - pseudoscalar & $\sigma_\text{obs}/\sigma_\text{theo}$ &  1.8 & 1.4 \\
 $g_\chi=1$                &                      &         &        \\
 $g_{\Pq}=1$                &                      &         &        \\
 $m_\PGc=1\GeV$                &                      &         &        \\ [\cmsTabSkip]
2HDM+\aboson                                 & $m_{\aboson}$     &  330\GeV  & 440\GeV  \\
 $\tan\beta=1$                &                      &         &        \\
 $m_\PGc=1\GeV$                &                      &         &        \\
 $\sin\theta=0.35$                &                      &         &        \\
$m_{\PH}=m_{\PSA}=1\TeV$  &                       &         &        \\[\cmsTabSkip]
2HDM+\aboson                                 & $m_{\PH}$     &  1200\GeV  & 1200\GeV  \\
 $\tan\beta=1$                &                      &         &        \\
 $m_\PGc=1\GeV$                &                      &         &        \\
 $\sin\theta=0.35$                &                      &         &        \\
$m_{\aboson}=100\GeV$  &                       &         &        \\ [\cmsTabSkip]
Invisible Higgs boson & $\mathcal{B}(\Ph \to {\text{invisible}})$   &   0.29 & 0.25 \\ [\cmsTabSkip]
ADD           & $M_{\mathrm{D}}$   & 2.8--2.9\TeV & 2.6--2.7\TeV \\
$n=2$--7    &                  &                    &                     \\ [\cmsTabSkip]
Unparticles                                      &  $\sigma$   &   0.26\unit{pb} &   0.24\unit{pb} \\ 
Scaling dimension $d_\textsf{U}$=1.5     &                    &                &                \\ 
\hline
  \end{tabular}
}
\end{table*}

\section{Summary}\label{sec:summary}
Events with a \PZ boson recoiling against missing transverse momentum in 
proton-proton collisions at the LHC are used to search for physics beyond 
the standard model. The results are interpreted in the context of several different models 
of the coupling mechanism between dark matter and ordinary matter: 
simplified models of dark matter with vector, axial-vector, scalar, and 
pseudoscalar mediators; invisible decays of a 125\GeV scalar Higgs boson; 
and a two-Higgs-doublet model with an extra pseudoscalar. Outside the context 
of dark  matter, models that invoke large extra dimensions or propose the production of unparticles could contribute to the same signature and are also considered. The 
observed limits on the production cross sections are used to constrain parameters 
of each of these models. The search utilizes a data set collected by the 
CMS experiment in 2016--2018, corresponding to an integrated luminosity of 137\fbinv 
at $\sqrt{s}=13\TeV$. No evidence of physics beyond the standard model is observed. 
Comparing to the previous results in this channel based on a partial data sample 
collected at $\sqrt{s}=13\TeV$ in 2016, corresponding to an integrated luminosity of 
approximately 36\fbinv for CMS~\cite{Sirunyan:2017qfc} and for ATLAS~\cite{Aaboud:2017bja}, the exclusion 
limits for simplified dark matter mediators, gravitons and unparticles are significantly extended. 
For the case of a 125\GeV scalar boson, an upper limit of 29\% is set 
for the branching fraction to fully invisible decays at 95\% confidence level. 
Results for the two-Higgs-doublet model with an additional pseudoscalar are presented in this final state  
and probe masses of the pseudoscalar mediator up to 440\GeV and of the heavy Higgs boson up to 
1200\GeV when the other model parameters are set to specific benchmark values.

\begin{acknowledgments}
  We congratulate our colleagues in the CERN accelerator departments for the excellent performance of the LHC and thank the technical and administrative staffs at CERN and at other CMS institutes for their contributions to the success of the CMS effort. In addition, we gratefully acknowledge the computing centers and personnel of the Worldwide LHC Computing Grid for delivering so effectively the computing infrastructure essential to our analyses. Finally, we acknowledge the enduring support for the construction and operation of the LHC and the CMS detector provided by the following funding agencies: BMBWF and FWF (Austria); FNRS and FWO (Belgium); CNPq, CAPES, FAPERJ, FAPERGS, and FAPESP (Brazil); MES (Bulgaria); CERN; CAS, MoST, and NSFC (China); COLCIENCIAS (Colombia); MSES and CSF (Croatia); RIF (Cyprus); SENESCYT (Ecuador); MoER, ERC IUT, PUT and ERDF (Estonia); Academy of Finland, MEC, and HIP (Finland); CEA and CNRS/IN2P3 (France); BMBF, DFG, and HGF (Germany); GSRT (Greece); NKFIA (Hungary); DAE and DST (India); IPM (Iran); SFI (Ireland); INFN (Italy); MSIP and NRF (Republic of Korea); MES (Latvia); LAS (Lithuania); MOE and UM (Malaysia); BUAP, CINVESTAV, CONACYT, LNS, SEP, and UASLP-FAI (Mexico); MOS (Montenegro); MBIE (New Zealand); PAEC (Pakistan); MSHE and NSC (Poland); FCT (Portugal); JINR (Dubna); MON, RosAtom, RAS, RFBR, and NRC KI (Russia); MESTD (Serbia); SEIDI, CPAN, PCTI, and FEDER (Spain); MOSTR (Sri Lanka); Swiss Funding Agencies (Switzerland); MST (Taipei); ThEPCenter, IPST, STAR, and NSTDA (Thailand); TUBITAK and TAEK (Turkey); NASU (Ukraine); STFC (United Kingdom); DOE and NSF (USA).

  \hyphenation{Rachada-pisek} Individuals have received support from the Marie-Curie program and the European Research Council and Horizon 2020 Grant, contract Nos.\ 675440, 752730, and 765710 (European Union); the Leventis Foundation; the A.P.\ Sloan Foundation; the Alexander von Humboldt Foundation; the Belgian Federal Science Policy Office; the Fonds pour la Formation \`a la Recherche dans l'Industrie et dans l'Agriculture (FRIA-Belgium); the Agentschap voor Innovatie door Wetenschap en Technologie (IWT-Belgium); the F.R.S.-FNRS and FWO (Belgium) under the ``Excellence of Science -- EOS" -- be.h project n.\ 30820817; the Beijing Municipal Science \& Technology Commission, No. Z191100007219010; the Ministry of Education, Youth and Sports (MEYS) of the Czech Republic; the Deutsche Forschungsgemeinschaft (DFG) under Germany's Excellence Strategy -- EXC 2121 ``Quantum Universe" -- 390833306; the Lend\"ulet (``Momentum") Program and the J\'anos Bolyai Research Scholarship of the Hungarian Academy of Sciences, the New National Excellence Program \'UNKP, the NKFIA research grants 123842, 123959, 124845, 124850, 125105, 128713, 128786, and 129058 (Hungary); the Council of Science and Industrial Research, India; the HOMING PLUS program of the Foundation for Polish Science, cofinanced from European Union, Regional Development Fund, the Mobility Plus program of the Ministry of Science and Higher Education, the National Science Center (Poland), contracts Harmonia 2014/14/M/ST2/00428, Opus 2014/13/B/ST2/02543, 2014/15/B/ST2/03998, and 2015/19/B/ST2/02861, Sonata-bis 2012/07/E/ST2/01406; the National Priorities Research Program by Qatar National Research Fund; the Ministry of Science and Higher Education, project no. 02.a03.21.0005 (Russia); the Programa Estatal de Fomento de la Investigaci{\'o}n Cient{\'i}fica y T{\'e}cnica de Excelencia Mar\'{\i}a de Maeztu, grant MDM-2015-0509 and the Programa Severo Ochoa del Principado de Asturias; the Thalis and Aristeia programs cofinanced by EU-ESF and the Greek NSRF; the Rachadapisek Sompot Fund for Postdoctoral Fellowship, Chulalongkorn University and the Chulalongkorn Academic into Its 2nd Century Project Advancement Project (Thailand); the Kavli Foundation; the Nvidia Corporation; the SuperMicro Corporation; the Welch Foundation, contract C-1845; and the Weston Havens Foundation (USA).
\end{acknowledgments}

\bibliography{auto_generated}
\cleardoublepage \appendix\section{The CMS Collaboration \label{app:collab}}\begin{sloppypar}\hyphenpenalty=5000\widowpenalty=500\clubpenalty=5000\vskip\cmsinstskip
\textbf{Yerevan Physics Institute, Yerevan, Armenia}\\*[0pt]
A.M.~Sirunyan$^{\textrm{\dag}}$, A.~Tumasyan
\vskip\cmsinstskip
\textbf{Institut f\"{u}r Hochenergiephysik, Wien, Austria}\\*[0pt]
W.~Adam, T.~Bergauer, M.~Dragicevic, J.~Er\"{o}, A.~Escalante~Del~Valle, R.~Fr\"{u}hwirth\cmsAuthorMark{1}, M.~Jeitler\cmsAuthorMark{1}, N.~Krammer, L.~Lechner, D.~Liko, T.~Madlener, I.~Mikulec, F.M.~Pitters, N.~Rad, J.~Schieck\cmsAuthorMark{1}, R.~Sch\"{o}fbeck, M.~Spanring, S.~Templ, W.~Waltenberger, C.-E.~Wulz\cmsAuthorMark{1}, M.~Zarucki
% ...
\vskip\cmsinstskip
\dag: Deceased\\
1:  Also at Vienna University of Technology, Vienna, Austria\\
% ...
\end{sloppypar}
\end{document}